\documentclass[journal,twoside]{IEEEtran}
\ifCLASSINFOpdf
\else
\fi

\usepackage{amsmath,amsthm}
\usepackage{enumerate}
\usepackage[noadjust]{cite}
\usepackage{}
\usepackage{graphicx,epstopdf}
\usepackage{xfrac,amssymb}
\usepackage[table]{xcolor}
\usepackage[normalem]{ulem}
\usepackage{flushend}
\usepackage{comment}

\graphicspath{{Figures/}} 

\usepackage{mdframed}

\usepackage{algorithm}
\usepackage[noend]{algpseudocode}
\algnewcommand{\Initialize}[1]{%
  \State \textbf{Initialize:}
  \Statex \hspace*{\algorithmicindent}\parbox[t]{.8\linewidth}{\raggedright #1}
}

\newcounter{MYtempeqncnt} 

\newcommand{\argmax}[1]{\underset{#1}{\operatorname{arg}\,\operatorname{max}}\;}
\newcommand{\argmin}[1]{\underset{#1}{\operatorname{arg}\,\operatorname{min}}\;}

\begin{document}
%
\title{Modified Vector Quantization for Small-Cell Access Point Placement with Inter-Cell Interference
}
%
%
%

\author{Govind~R.~Gopal,~\IEEEmembership{Student~Member,~IEEE,}~Elina~Nayebi,~\IEEEmembership{Member,~IEEE,}~\\~Gabriel~Porto~Villardi,~\IEEEmembership{Senior~{Member},~IEEE,}~and~Bhaskar~D.~Rao,~\IEEEmembership{Fellow,~IEEE}
\thanks{G. R. Gopal and B. D. Rao are with the Department
of Electrical and Computer Engineering, University of California, San Diego, La Jolla,
CA 92093 USA (e-mail: ggopal@ucsd.edu; brao@ucsd.edu).
The work of G. R. Gopal was supported in part by National Science Foundation (NSF) Grant No. CCF-1617365 and the Center for Wireless Communications (CWC), University of California, San Diego.
}
\thanks{E. Nayebi is with Apple Inc., Cupertino, CA 95014 USA (e-mail: e.nayebi@gmail.com). The work of E. Nayebi was carried out when she was a PhD student at the University of California, San Diego.
}
\thanks{G. P. Villardi is with the Wireless Network Research Center of the National Institute of Information and Communications Technology (NICT), Yokosuka 239-0847, Japan. This work was done when he was a visiting scholar with the Qualcomm Institute of Calit2, 
University of California, San Diego, La Jolla, CA 92093 USA.  (e-mail: gpvillardi@nict.go.jp).}
}

\maketitle
\begin{abstract}

In this paper, we explore the small-cell uplink access point (AP) placement problem in the context of throughput-optimality and provide solutions while taking into consideration inter-cell interference. First, we briefly review the vector quantization (VQ) approach and related single user throughput-optimal formulations for AP placement. Then, we investigate the small-cell case with multiple users and expose the limitations of mean squared error based VQ for solving this problem. While the Lloyd algorithm from the VQ approach is found not to strictly solve the small-cell case, based on the tractability and quality of resulting AP placement, we deem it suitable as a simple and appropriate framework to solve more complicated problems. Accordingly, to minimize ICI and consequently enhance achievable throughput, we design two Lloyd-type algorithms, namely, the Interference Lloyd algorithm and the Inter-AP Lloyd algorithm, both of which incorporate ICI in their distortion functions. Simulation results show that both of the proposed algorithms provide superior 95\%-likely rate over the traditional Lloyd algorithm and the Inter-AP Lloyd algorithm yields a significant increase of up to 36.34\% in achievable rate over the Lloyd algorithm.

\end{abstract}


%
\IEEEpeerreviewmaketitle

\section{Introduction}
\label{sec:intro}
\IEEEPARstart{T}{he} past decade has witnessed the surge of wireless communications technologies that have significantly raised the bits-per-second-per-hertz figure of merit and throughput of wireless networks in order to cope with the ongoing widespread adoption of mobile broadband by society. One of these key technologies, massive multiple-input-multiple-output (massive MIMO) \cite{mar10,lar14,and14,llu14,vil19}, utilizes hundreds of antenna elements and is particularly appealing since intra-cell interference and small-scale fading are naturally canceled out due to favorable propagation and channel hardening \cite{Chen2018}, leading to low-cost hardware implementations \cite{Emil14}. Distributed antenna systems (DASs), especially in the form of distributed MIMO \cite{hua11,tru13,rog14,ngo17,wan09,par12,yan15}, give rise to even higher average rates over co-located MIMO systems \cite{cho07,wan12,tru13}. In general, distributed massive MIMO can either be cooperative or non-cooperative, with cell-free massive MIMO \cite{Nayebi2016,nay17} being an example of the former. Although cooperation between the distributed antenna elements in the cell-free approach assists in mitigating interference between users and further increases spectral efficiency over non-cooperative massive MIMO systems, the required user-related information exchange occupies a significant portion of the usually limited back-haul capacity of wireless systems \cite{ges10,hua11,irm11}. As a result, despite the benefits of cell-free massive MIMO systems, near future deployments of 5G (3GPP Rel-15 and Rel-16 \cite{Ghosh2019}) and WiFi 6 (IEEE 802.11ax \cite{Khorov2019}) will still be based on the concept of small-cells, possibly leaving cell-free approaches to posterior deployments of the technology. Hence, controlling inter-cell interference (ICI) continues to be a major system design problem, which will actually assume much greater significance with the expected network densification of Beyond 5G systems. Currently, ICI is dealt with in the standards by advanced scheduling techniques such as basic service set (BSS) coloring \cite{Khorov2019} and dynamic time division duplex (D-TDD) \cite{Ding2018}. Also, controlling ICI is of great importance to public protection and disaster relief (PPDR) wireless networks occupying the 700 MHz (and below) frequency bands due to their desirable propagation characteristics and higher signal penetration capabilities, which can cause severe service outages to adjacent emergency networks even in not-so-dense deployments \cite{Vil2012,Vil2012_2,Vil2011,Grosspietsch2013}.

In this work, we exploit another degree of freedom in system design, namely access point (AP) placement, in order to tackle ICI in  small-cell wireless  systems with non-uniform user distributions.  
To contextualize the discussion, consider a large gathering such as a sporting event, where sections in the stadium see a different number and arrangement of spectators depending on the crowd on the day of the event. To avoid service interruption, more APs should be placed where the number of spectators is larger, and vice versa, leading to the concept of smart stadiums. 
Additionally, flexible AP deployment is of utmost importance in the infrequent emergency and disaster relief situations, where deployments should be tailored to the time-specific coverage and service requirements, therefore following the dynamics of the emergency event \cite{Vil2012}.
Thus, the question of interest is: \emph{How do we optimally place the APs given the distribution of users?}
In recent times, the AP (or antenna) placement problem has attracted a great deal of attention \cite{wan09,par12,yan15,abr18,kar19}, however, optimizing the AP or antenna locations by maximizing a signal-to-noise ratio (SNR) objective function alone has traditionally been the standard approach. The authors of \cite{wan09} consider a DAS and optimize the cell averaged ergodic capacity based only on SNR and neglect ICI. Using the square distance criterion, they notice similarities with codebook design in vector quantization (VQ), which enables the utilization of the well-known (for ease of implementation) Lloyd algorithm to solve the antenna placement problem. In \cite{yan15}, the average achievable per-user rate of uniformly distributed users is optimized in order to find the radius of a circular antenna array; however, due to the adoption of a single-cell model, no ICI is considered. Circular antenna array deployments based on average rate optimization are also considered in \cite{par12} based on one-cell and two-cell models, with the latter model accounting for leakage interference alone. Additionally, the authors of \cite{abr18} simulated an indoor wireless environment where they generated a 10-fold improvement in the distributed system capacity over the co-located one. Further, placing APs in accordance with the user densities generated a significant increase (40\% over uniform AP placement) in system capacity. 
Minimization of the total power consumption of a heterogeneous wireless sensor network with APs (first tier) and fusion centers (second tier) in \cite{kar19} results in a two-tier Lloyd-type algorithm to place both APs and fusion centers, but without addressing ICI.
Recently, unmanned aerial vehicles (UAVs) equipped with base stations have also been considered in the context of AP placement \cite{gal16,bor16,lyu17,zha18,lai19,xie19,guo20,zen19}.

In the abovementioned works, the suitability of the Lloyd algorithm from VQ to throughput-optimal AP placement has not been investigated. VQ considers only a single user, and the objective function is averaged over the position of this user. This approach, however, does not conform with the small-cell scenario where there are multiple users, one from each cell, communicating with the serving AP in its cell. Further, ICI has been neglected, therefore leading to AP placements that yield sub-optimal throughput.
Hence, in this work, we device a non-cooperative small-cell system based on the Lloyd algorithm, which we show can solve for near-optimal AP locations, in terms of the fundamental performance measure of throughput, while considering ICI.

\subsection*{Contributions}
To the best of our knowledge, solutions to the AP placement problem based on the Lloyd algorithm and that are derived from a detailed analysis of throughput optimality, while incorporating ICI, have not been provided in the literature. Hence, in this work on small-cell AP placement, our contributions are as follows.
\begin{itemize}
	\item We first formulate various single user AP placement problems for throughput optimality in terms of rate, SNR, and a higher exponent for the user-AP distance (as opposed to squared distance). We explore the relationship of the Lloyd algorithm from VQ to these problems. We then study the multiple user case and address the small-cell AP placement problem. Although our analysis determines that the application of the Lloyd algorithm to small-cell AP placement is not ideal, we find that the Lloyd algorithm, apart from being easy to implement, is quite effective in solving the placement problem as a baseline algorithm and yields near-optimal AP locations.
	\item We present two methods to incorporate ICI into the optimization function of the Lloyd algorithm. Consequently, the distortion function of the Lloyd algorithm is modified and two Lloyd-type algorithms for AP placement that are aware of ICI and, as a result, maximize achievable per-user SINR, are proposed, namely, the Interference Lloyd algorithm and Inter-AP Lloyd algorithm. 
\end{itemize}

The remainder of this paper is organized as follows. Section \ref{sec:system_model} outlines the small-cell model used throughout the paper. VQ and the application of the Lloyd algorithm to the AP placement problem is described in Section \ref{sec:Small-Cell AP Placement Problem Formulation}. Mathematical formulations of throughput optimality for single user and multiple user cases are provided in \ref{sec:Problem Formulations}. The ensuing section \ref{sec:acct_for_ici} presents the formulations and solutions for including ICI in the VQ approach. Cell association strategies for each of the proposed algorithms are elucidated in Section \ref{sec:Cell Association}. The simulation methodology and results are stated in Section \ref{sec:Simulation Methodology}. Finally, we provide concluding remarks in Section \ref{sec:Conclusion}.


\section{System Model}
\label{sec:system_model}

We use the small-cell model detailed in \cite{nay18a} and \cite[Ch. 4]{nay18t}, which is reproduced here for completeness.
%
%
Also, throughout this paper we use bold symbols to denote vectors, $\mathbb{E}\{\cdot\}$ is the expectation operator, $||\cdot||$ represents the $\ell_2$-norm of a vector, and all logarithms are to the base 2.
Now, consider a geographical area where $K$ single-antenna users are distributed, according to some probability density function (pdf) $f_{\mathbf{P}}(\mathbf{p})$, where $\mathbf{p} \in \mathbb{R}^2$ is the random vector denoting the position of a user. There are $M$ single-antenna APs that serve the users in this area. The location of an AP is denoted by $\mathbf{q} \in \mathbb{R}^2$. 
All APs are connected via error-free backhaul links to the network controller\footnote{The NC is where the proposed placement algorithms to be described in detail in the remainder of this manuscript will be loaded and executed.} (NC), so that it knows the positions of the APs and their respective users. 
For simplicity, a narrowband flat-fading channel is considered. With $m = 1,2,\ldots,M$ and $k = 1,2,\ldots,K$, the channel coefficient between the $m^{\text{th}}$ AP and $k^{\text{th}}$ user is
\begin{equation}
g_{mk} = \sqrt{\beta_{mk}}h_{mk},
\end{equation}
where $\beta_{mk}$ and $h_{mk}\sim\mathcal{CN}(0,1)$ are the large-scale and small-scale fading coefficients, respectively. $h_{mk}$ is assumed to remain constant during a coherent interval and change independently in the next, and is independent of $\beta_{mk}$. 
The large-scale fading coefficients are modeled as 
\begin{equation}
\beta_{mk} = \left\{
\begin{array}{cl}
c_0, & \left|\left| \mathbf{p}_k-\mathbf{q}_m \right|\right| \leq r_0, \\
\frac{ c_1 z_{mk} }{ \left|\left| \mathbf{p}_k-\mathbf{q}_m \right|\right|^{\gamma} }, & \left|\left| \mathbf{p}_k-\mathbf{q}_m \right|\right| > r_0, \\
\end{array} 
\right.
\label{eqn:pathloss}
\end{equation}
where $\mathbf{p}_k \in \mathbb{R}^2$ and $\mathbf{q}_m \in \mathbb{R}^2$ represent the locations of the $k^{\text{th}}$ user and $m^{\text{th}}$ AP, respectively. Here, $\gamma$ is the pathloss exponent, $z_{mk}$ is the log-normal shadow fading coefficient, and $c_0$, $c_1$, and $r_0$ are constants. These coefficients can also be estimated by either ray-tracing \cite{yun15} or data-driven \cite{ost10} approaches.

The uplink transmission model used in this work schedules users in a round robin fashion with their serving APs using time-division multiple access (TDMA). Thus, each AP serves only one user in a time slot.
In the small-cell set-up, each of the $M$ cells corresponds to each of the $M$ APs, and pursuant with the uplink model, the user in each cell communicating with its associated AP causes interference to all other APs. Now, letting $k_m$ denote a user in the cell associated with AP $m$, the received signal $y_m$ at this AP is
\begin{equation}
y_m = \sum\limits_{m^{\prime}=1}^M \sqrt{\rho_r}g_{mk_{m^{\prime}}}s_{k_{m^{\prime}}} + w_m,
\label{eqn:receivedsignal}
\end{equation}
where $\rho_r$ is the uplink transmit power, $s_{k_m}$ is the data symbol with $\mathbb{E}\{|s_{k_m}|^2\}=1$ (unit power), and $w_m \sim \mathcal{CN}(0,1)$ is the additive noise. A matched filter (MF) employed at the AP $m$ estimates the data symbol $s_{k_m}$ of user $k_m$ as
\begin{equation}
\begin{aligned}
\hat{s}_{k_m} &= \frac{ g^*_{mk_m}  }{ |g_{mk_m}| } y_m,\\
&= \underbrace{\sqrt{\rho_r}|g_{mk_m}| s_{k_m}}_{T_{\text{des}}\text{: desired term}} + \underbrace{\sum\limits_{\substack{m^{\prime}=1\\m^{\prime}\neq m}}^M 
	\sqrt{\rho_r} \frac{ g^*_{mk_m}  }{ |g_{mk_m}| } g_{mk_{m^{\prime}}} s_{k_{m^{\prime}}}}_{T_{\text{int}}\text{: interference term}} + v_m,
\end{aligned}
\label{eqn:est_symb}
\end{equation}
where $v_m \sim \mathcal{CN}(0,1)$. Considering $T_{\text{int}}$ in \eqref{eqn:est_symb} as noise, the signal-to-interference-plus-noise ratio (SINR) achieved by user $k_m$ at AP $m$ is derived to be
\begin{equation}
\phi_{k_m} = \frac{ \rho_r \beta_{mk_m} |h_{mk_m}|^2  }{ 1 + \rho_r \sum\limits_{\substack{m^{\prime}=1\\m^{\prime}\neq m}}^M \beta_{mk_{m^{\prime}}} |h_{mk_{m^{\prime}}}|^2 }.
\label{eqn:sinr}
\end{equation}

\section{Vector Quantization and AP Placement}
\label{sec:Small-Cell AP Placement Problem Formulation}

In this section, we provide an overview of VQ and how the Lloyd algorithm is currently used in its basic form, to solve for AP placement. Note that Section \ref{sec:Problem Formulations} will investigate the suitability of the Lloyd algorithm to obtain AP locations.

\subsection{Overview of Vector Quantization}

In VQ, the random vector to be quantized is $\mathbf{x} \in \mathbb{R}^p$, where $p$ is the dimension, and the two main steps to be designed are the encoding and decoding steps. 
The encoder $\mathcal{E}$ splits the domain under consideration into $N$ regions (called Voronoi regions, each corresponding to a bit sequence of length $\log_2 N$) and assigns a region $\mathcal{R}$ to the input vector $\mathbf{x}$. The encoder performs the following mapping
\begin{equation}
	\mathcal{E}: \mathbb{R}^p \rightarrow \{\mathcal{R}_1,\mathcal{R}_2,\ldots,\mathcal{R}_N\}.
\end{equation}
The decoder $\mathcal{D}$ then assigns to each region $\mathcal{R}_n$, where $n = 1,2,\ldots,N$, a codepoint $\hat{\mathbf{x}}_n$, and performs the mapping
\begin{equation}
	\mathcal{D}: \{\mathcal{R}_1,\mathcal{R}_2,\ldots,\mathcal{R}_N\} \rightarrow \{\hat{\mathbf{x}}_1,\hat{\mathbf{x}}_2,\ldots,\hat{\mathbf{x}}_N\}.
\end{equation}
The set of codepoints $\{\hat{\mathbf{x}}_1,\hat{\mathbf{x}}_2,\ldots,\hat{\mathbf{x}}_N\}$ is collectively the codebook.
Thus, the quantizer $\mathcal{Q}$ assigns for every input $\mathbf{x}$, one of $N$ codepoints, and is given as
\begin{equation}
\mathcal{Q}(\mathbf{x}) = \mathcal{D}(\mathcal{E}(\mathbf{x})) = \hat{\mathbf{x}}_{\mathcal{E}(\mathbf{x})},
\end{equation}
where $\hat{\mathbf{x}}_{\mathcal{E}(\mathbf{x})}$ specifies that the output codepoint is a function of the input vector and for simplicity in notation, we assume that $\mathcal{E}(\mathbf{x})$ denotes the index of the region that is specifies. 
The encoder $\mathcal{E}$ assigns to the input $\mathbf{x}$, the region that is closest to it, defined in terms of a distortion function $d$ between the input vector and a codepoint. The codepoint corresponding to the region can formally be written as
\begin{equation}
    \hat{\mathbf{x}}_{\mathcal{E}(\mathbf{x})} = \argmin{\hat{\mathbf{x}}_n} d(\mathbf{x},\hat{\mathbf{x}}_n).
    \label{eqn:VQ_arg_min_distortion}
\end{equation}
Taking the average of the distortion function over the distribution of the input vector, the optimization problem of VQ is written as
\begin{equation}
    \argmin{\hat{\mathbf{x}}_1,\hat{\mathbf{x}}_2,\ldots,\hat{\mathbf{x}}_N} \mathbb{E}_{\mathbf{x}}\left\{ d(\mathbf{x},\hat{\mathbf{x}}_{\mathcal{E}(\mathbf{x})}) \right\}.
    \label{eqn:VQ_opt_problem}
\end{equation}

To solve the optimization of \eqref{eqn:VQ_opt_problem}, the goal is to find the optimal encoder and decoder jointly, which is difficult. Hence, it is split into two tasks, which are to find a optimal encoder given a fixed decoder and a optimal decoder given a fixed encoder, and form the two necessary conditions for quantizer optimality. The main methodology then is to alternate between these two tasks in order to converge to a reasonable solution. Accordingly, finding the best encoder given the decoder involves determining the best regions given fixed codepoints. This leads to the \textit{Nearest Neighbor Condition (NNC)}
\begin{equation}
\mathcal{R}_n = \{ \mathbf{x} : d(\mathbf{x},\hat{\mathbf{x}}_n) \leq d(\mathbf{x},\hat{\mathbf{x}}_l), \forall l \neq n \}.
\end{equation}
Next, finding the best decoder given the encoder involves determining the best codepoints given the regions. This is the \textit{Centroid Condition (CC)}, given by
\begin{equation}
\hat{\mathbf{x}}_n = \mathrm{Cent} \{ \mathbf{x} | \mathbf{x} \in \mathcal{R}_n \},
\end{equation}
where the centroid $\mathrm{Cent}$\footnote{The centroid is defined \cite{ger91} as \begin{equation*}
    \mathrm{Cent} \{ \mathbf{x} | \mathbf{x} \in \mathcal{R}_n \} = \argmin{\hat{\mathbf{x}}_n} \mathbb{E} \{ d(\mathbf{x},\hat{\mathbf{x}}_n) | \mathbf{x} \in \mathcal{R}_n \}.
\end{equation*}} of region $R_n$ gives the codepoint $\hat{\mathbf{x}}_n$ for the region. The Lloyd algorithm alternates between the NNC and CC steps until convergence and yields the optimal codepoints.

\subsection{The Lloyd Algorithm for AP Placement}

If the VQ approach were to be used to solve for small-cell AP placement, then the random vector to be quantized is the 2-D position $\mathbf{p}$ of a \emph{single} user. The Voronoi regions are the cells $\mathcal{C}_m$ and the codepoints are the AP locations $\mathbf{q}_m$, where $m = 1,2,\ldots, M$.
The optimization problem in 
\eqref{eqn:VQ_opt_problem} can be written by using similar notations and taking the average over the user positions, as follows
\begin{equation}
    \argmin{\mathbf{q}_1,\mathbf{q}_2,\ldots,\mathbf{q}_M} \mathbb{E}_{\mathbf{p}}\left\{ d(\mathbf{p},\mathbf{q}_{\mathcal{E}(\mathbf{p})}) \right\}.
\end{equation}
It is worth reiterating that $\mathcal{E}(\mathbf{p})$ indexes the nearest AP that the user at $\mathbf{p}$ associates to. 
The objective function above can be written as
\begin{equation}
\begin{aligned}
J_{\text{VQ}} &= \mathbb{E}_{\mathbf{p}}\left\{ d(\mathbf{p},\mathbf{q}_{\mathcal{E}(\mathbf{p})}) \right\},\\
&=  \int\limits_{-\infty}^{\infty} d(\mathbf{p},\mathbf{q}_{\mathcal{E}(\mathbf{p})}) f_{\mathbf{P}}(\mathbf{p}) \mathrm{d}\mathbf{p},\\
&= \sum\limits_{m=1}^M \left[\int\limits_{\mathbf{p} \in \mathcal{C}_m} d(\mathbf{p},\mathbf{q}_m) f_{\mathbf{P}}(\mathbf{p} | \mathbf{p} \in \mathcal{C}_m) \mathrm{d}\mathbf{p}\right] \Pr(\mathbf{p} \in \mathcal{C}_m),\\
&= \sum\limits_{m=1}^M S_m \Pr(\mathbf{p} \in \mathcal{C}_m),
\end{aligned}
\label{eqn:VQ_opt_func}
\end{equation}%
where the penultimate step arises by splitting the integral in the previous step into the cells (Voronoi regions) with their respective codepoints and the quantity $S_m$ is defined as
\begin{equation}
    S_m = \int\limits_{\mathbf{p} \in \mathcal{C}_m} d(\mathbf{p},\mathbf{q}_m) f_{\mathbf{P}}(\mathbf{p} | \mathbf{p} \in \mathcal{C}_m) \mathrm{d}\mathbf{p}.
    \label{eqn:Sm_definition}
\end{equation}

To solve for the optimal AP locations, the most often used distortion function is the squared Euclidean distance \begin{equation}
d_{\text{SE}}(\mathbf{p},\mathbf{q}_{\mathcal{E}(\mathbf{p})}) = \left|\left| \mathbf{p}-\mathbf{q}_{\mathcal{E}(\mathbf{p})} \right|\right|^2,
\label{eqn:sq_error_distortion}
\end{equation}
and the objective function in \eqref{eqn:VQ_opt_func} then becomes the mean squared error (MSE). In this paper, we retain the name `Lloyd algorithm' for the algorithm that solves \eqref{eqn:VQ_opt_func} using $d_{\text{SE}}$, the steps of which are provided in Algorithm \ref{alg:lloyd}. For all other distortion functions, we will use the name `Lloyd-type algorithm'.
\begin{algorithm}
\caption{Lloyd Algorithm With Squared Error Distortion}\label{alg:lloyd}
\begin{algorithmic}[1]
\State Initialize random AP locations $\mathbf{q}_1^{(0)},\mathbf{q}_2^{(0)},\ldots,\mathbf{q}_M^{(0)}$.
\State Use the NNC to determine the cells $\mathcal{C}_1^{(i+1)},\mathcal{C}_2^{(i+1)},\ldots,\mathcal{C}_M^{(i+1)}$ such that
\begin{equation*}
\mathcal{C}_m^{(i+1)} \!=\! \left\{\! \mathbf{p}_k \!:\! d_{\text{SE}}\!\left(\!\mathbf{p}_k,\mathbf{q}_m^{(i)}\!\right) \!\leq \! d_{\text{SE}}\!\left(\!\mathbf{p}_k,\mathbf{q}_l^{(i)}\!\right)\!, \forall l \!\neq m \!\right\}.
\label{eqn:nnc_lloyd}
\end{equation*}
\State Use the CC to determine the AP locations $\mathbf{q}_1^{(i+1)},\mathbf{q}_2^{(i+1)},\ldots,\mathbf{q}_M^{(i+1)}$ such that
\begin{equation*}
\mathbf{q}_m^{(i+1)} = \sum\limits_{\mathbf{p}_k \in \mathcal{C}_m^{(i+1)}} \mathbf{p}_k.
\end{equation*}
\State Repeat from step 2 until convergence (MSE falls below a threshold).
\end{algorithmic}
\end{algorithm}
Note that when the Lloyd algorithm is implemented, we use the $K$ realization of users at positions $\mathbf{p}_k$, $k = 1,2,\ldots,K$, as described in Section \ref{sec:system_model}. We will use this notation for all the Lloyd-type algorithms that follow. Also, observe that in the CC step in \ref{alg:lloyd}, as a result of $d_{\text{SE}}$, the centroid is replaced by the expectation which is evaluated by using the sample average over the user positions $\mathbf{p}_k$ present in cell $\mathcal{C}_m$.

An interesting observation here is that the VQ framework presented above considers only the positions of the users and APs, and hence is independent of both the small-scale fading and shadow fading components of the wireless system since these quantities are not dependent on the user and AP positions. These random quantities thus do not play a role in AP placement using VQ. It is also very important to note here that VQ considers only a single user to be quantized and the average over the distribution of that user is taken. However, this does not conform to our small-cell system model, where $M$ users are each communicating with its serving AP at the same time. Hence, the VQ approach does not strictly solve the small-cell AP placement problem.

\section{Throughput Formulations and Solutions Without Inter-Cell Interference}
\label{sec:Problem Formulations}

In this section, we will describe throughput optimization via various formulations, such as average rate and SNR, and provide solutions to obtain optimal AP locations. We start by considering the single user scenario inherent to VQ and expand to a more realistic one in which multiple users are present. We also illustrate, by formulation only, the case where ICI is present. In summary, we argue how the Lloyd algorithm, despite its simplicity, is suitable for small-cell AP placement.

\subsection{Single User Case}

\subsubsection{Rate}
\label{sssec:single_user_rate}
The single user case is the simplest case wherein a user at location $\mathbf{p}$ alone is considered. 
We start our analysis with per-user rate, which is the common measure of interest, achieved by a user at $\mathbf{p}$ with its nearest AP at $\mathbf{q}_{\mathcal{E}(\mathbf{p})}$, as per the VQ principles discussed above. We also approximate the large-scale fading coefficients, given in \eqref{eqn:pathloss}, by
\begin{equation}
\beta_{\mathcal{E}(\mathbf{p})} \approx \frac{ c_1 z_{\mathcal{E}(\mathbf{p})} }{ \left|\left| \mathbf{p}-\mathbf{q}_{\mathcal{E}(\mathbf{p})} \right|\right|^{\gamma} },
\label{eqn:approx_beta}
\end{equation}
since $r_0$ is much smaller than the dimensions of the area under consideration. Note that the second subscript has been dropped for the ensuing analyses, since we consider a single user. 
Let us define the average rate, utilizing the SNR $\psi_{k_{\mathcal{E}(\mathbf{p})}}$ (obtained from \eqref{eqn:sinr} by neglecting ICI and replacing $m$ with $\mathcal{E}(\mathbf{p})$) as follows
\begin{equation}
    \overline{r}(\underline{\mathbf{q}}) = \mathbb{E}_{\mathcal{A},\mathbf{p}} \left\{ \log\left(1 + \psi_{k_{\mathcal{E}(\mathbf{p})}}\right) \right\},
    \label{eqn:rate_snr}
\end{equation}
where we average over the user position $\mathbf{p}$, the random quantities $h_{\mathcal{E}(\mathbf{p})}$ and $z_{\mathcal{E}(\mathbf{p})}$, $\mathcal{A} = \{h_{\mathcal{E}(\mathbf{p})},z_{\mathcal{E}(\mathbf{p})}\}$ for brevity, and we use the notation $\underline{\mathbf{q}} = \{\mathbf{q}_1,\mathbf{q}_2,\ldots,\mathbf{q}_M\}$ to show that the average rate is a function of the $M$ AP locations alone.
Similar to VQ in the previous section, we average out the small-scale and shadow fading components defined in $\mathcal{A}$ since they are position independent and do not contribute to the optimal placement of APs.
Assuming high SNR $(\psi_{k_{\mathcal{E}(\mathbf{p})}} \gg 1)$, we can write \eqref{eqn:rate_snr} as
\begin{equation}
    \overline{r}(\underline{\mathbf{q}}) = \mathbb{E}_{\mathcal{A},\mathbf{p}} \left\{ \log\left(\frac{ \rho_r c_1 |h_{\mathcal{E}(\mathbf{p})}|^2 z_{\mathcal{E}(\mathbf{p})} }{ \left( \left|\left| \mathbf{p}-\mathbf{q}_{\mathcal{E}(\mathbf{p})} \right|\right|^2\right)^{\frac{\gamma}{2}}} \right) \right\}, 
    \label{eqn:rate_high_snr}
\end{equation}
and we wish to perform the optimization
\begin{equation}
    \argmax{\mathbf{q}_1,\mathbf{q}_2,\ldots,\mathbf{q}_M} \overline{r}(\underline{\mathbf{q}}).
\end{equation}
After averaging and removing the terms that are not involved in the optimization in \eqref{eqn:rate_high_snr}, we obtain
\begin{equation}
    \argmin{\mathbf{q}_1,\mathbf{q}_2,\ldots,\mathbf{q}_M} \mathbb{E}_{\mathbf{p}} \left\{ \log\left(\left|\left| \mathbf{p}-\mathbf{q}_{\mathcal{E}(\mathbf{p})} \right|\right|^2 + \epsilon\right) \right\},
    \label{eqn:rate_mm_opt}
\end{equation}
where we have added a constant $\epsilon > 0$ (typically very small) to prevent the logarithm from approaching negative infinity if the user position $\mathbf{p}$ were to overlap with the position of the nearest AP $\mathbf{q}_{\mathcal{E}(\mathbf{p})}$. Note that $\epsilon$ could correspond to the pathloss at a reference distance or even height of the AP.
The objective function to be optimized above is concave as a result of which the Majorization-Minimization (MM) technique \cite{ort70} can be used to acquire a solution to the centroid computation (CC) step. The NNC steps is is the same as before. The MM technique upperbounds the objective function by a surrogate function and minimizes the surrogate through an iterative method. Solving the objective function in \eqref{eqn:rate_mm_opt} using the MM method results in an iterative solution with the following two update equations
\begin{equation}
    \begin{aligned}
    \mathbf{q}_m^{(j+1)} &= \frac{ \sum\limits_{\mathbf{p}_k \in \mathcal{C}_m} w_k^{(j)} \mathbf{p}_k  }{ \sum\limits_{\mathbf{p}_k \in \mathcal{C}_m} w_k^{(j)} },\\
			w_k^{(j+1)} &= \frac{ 1 }{||\mathbf{q}_m^{(j+1)}-\mathbf{p}_k||^2 + \epsilon}, \ \forall \mathbf{p}_k \in \mathcal{C}_m,
    \end{aligned}
    \label{eqn:mm_iter_soln}
\end{equation}
where $j$ denotes the MM iteration index.
%


In summary, to solve for the AP locations, we can now formulate a Lloyd-type algorithm with the NNC step remaining the same as that in the Lloyd algorithm, i.e., with $d_{\text{SE}}$, and the CC step replaced by the above iterative solution of \eqref{eqn:mm_iter_soln}.
We call this Lloyd-type algorithm as the \emph{MM-Lloyd algorithm}. The proof of \eqref{eqn:mm_iter_soln} is left to Appendix \ref{app:mm_proof} and the algorithm is provided in Algorithm \ref{alg:mm_lloyd}.
\begin{algorithm}
\caption{MM-Lloyd Algorithm}\label{alg:mm_lloyd}
\begin{algorithmic}[1]
\State Initialize random AP locations $\mathbf{q}_1^{(0)},\mathbf{q}_2^{(0)},\ldots,\mathbf{q}_M^{(0)}$.
\State  Use the NNC to determine the cells $\mathcal{C}_1^{(i+1)},\mathcal{C}_2^{(i+1)},\ldots,\mathcal{C}_M^{(i+1)}$ such that
\begin{equation*}
\mathcal{C}_m^{(i+1)} \!=\! \left\{\! \mathbf{p}_k \!:\! d_{\text{SE}}\!\left(\!\mathbf{p}_k,\mathbf{q}_m^{(i)}\!\right) \!\leq \! d_{\text{SE}}\!\left(\!\mathbf{p}_k,\mathbf{q}_l^{(i)}\!\right)\!, \forall l \!\neq m \!\right\}.
\end{equation*}
\State Use MM iterations to determine the AP locations $\mathbf{q}_1^{(i+1)},\mathbf{q}_2^{(i+1)},\ldots,\mathbf{q}_M^{(i+1)}$ with the update equations
\begin{equation*}
    \begin{aligned}
    \mathbf{q}_m^{(j+1)} &= \frac{ \sum\limits_{\mathbf{p}_k \in \mathcal{C}_m^{(i+1)}} w_k^{(j)} \mathbf{p}_k  }{ \sum\limits_{\mathbf{p}_k \in \mathcal{C}_m^{(i+1)}} w_k^{(j)} },\\
			w_k^{(j+1)} &= \frac{1}{||\mathbf{q}_m - \mathbf{p}_k||^2 + \epsilon}, \ \forall \mathbf{p}_k \in \mathcal{C}_m^{(i+1)},
    \end{aligned}
\end{equation*}
where $\mathbf{q}_m^{(i+1)} = \mathbf{q}_m^{(j+1)}$ after convergence.
\State Repeat from step 2 until convergence.
\end{algorithmic}
\end{algorithm}

\subsubsection{SNR}

In contrast to rate maximization, if throughput is measured by SNR averaged over the user location $\mathbf{p}$, then we can show that the simple case of SNR maximization is equivalent to the VQ optimization problem given in \eqref{eqn:VQ_opt_func}. 
Let us write the average achievable SNR as
\begin{equation}
    \overline{\psi}(\underline{\mathbf{q}}) = \mathbb{E}_{\mathcal{A},\mathbf{p}} \left\{ \frac{ \rho_r c_1 |h_{\mathcal{E}(\mathbf{p})}|^2 z_{\mathcal{E}(\mathbf{p})} }{ \left|\left| \mathbf{p}-\mathbf{q}_{\mathcal{E}(\mathbf{p})} \right|\right|^{\gamma}} \right\},
\end{equation}
which is lower bounded by applying Jensen's inequality as
\begin{equation}
    \overline{\psi}(\underline{\mathbf{q}}) \geq \mathbb{E}_{\mathcal{A}} \left\{ \frac{ \rho_r c_1 |h_{\mathcal{E}(\mathbf{p})}|^2 z_{\mathcal{E}(\mathbf{p})} }{\left(\mathbb{E}_{\mathbf{p}}\left\{ \left|\left| \mathbf{p}-\mathbf{q}_{\mathcal{E}(\mathbf{p})} \right|\right|^{2}\right\}\right)^{\frac{\gamma}{2}}} \right\}.
\end{equation}
Maximizing $\overline{\psi}(\underline{\mathbf{q}})$ to obtain the AP locations is the same as minimizing the term in the denominator, leading to the same objective function \eqref{eqn:VQ_opt_func} in VQ. The optimization problem is
\begin{equation}
    \argmin{\mathbf{q}_1,\mathbf{q}_2,\ldots,\mathbf{q}_M} \mathbb{E}_{\mathbf{p}}\left\{ \left|\left| \mathbf{p}-\mathbf{q}_{\mathcal{E}(\mathbf{p})} \right|\right|^{2}\right\}.
    \label{eqn:snr_argmin}
\end{equation}
As before, this is solved using the Lloyd algorithm with $d_{\text{SE}}$ (Algorithm \ref{alg:lloyd}).
For consistency in future discussions, we introduce the notation $d(\mathbf{p},\underline{\mathbf{q}})$ as a general form of distortion measure with $\underline{\mathbf{q}} = \{\mathbf{q}_1,\mathbf{q}_2,\ldots,\mathbf{q}_M\}$. Hence, the squared error distortion function in \eqref{eqn:sq_error_distortion} is written in the general form as
\begin{equation}
    d_{\text{SE}}\left(\mathbf{p},\underline{\mathbf{q}}\right) = \left|\left| \mathbf{p}-\mathbf{q}_{\mathcal{E}(\mathbf{p})} \right|\right|^{2}.
    \label{eqn:MSE_distortion}
\end{equation}

\subsubsection{Higher exponent for user-AP distance}

The objective function in the Lloyd algorithm is proportional to the square of the user-AP distance while that in the MM-Lloyd algorithm is proportional to the logarithm of the squared distance. This means that the MM-Lloyd algorithm disproportionately considers the contribution of users, as the logarithm suppresses the larger distances inherent to users at the cell borders, in comparison to the Lloyd algorithm. To overcome this effect, we can design another optimization function that exponentially scales up large distances relative to the Lloyd algorithm by raising the distance to a higher power. This higher exponent $\chi > 2$ also characterizes higher frequency (e.g., mmWave) communications. The optimization problem can then be represented as
\begin{equation}
    \argmin{\mathbf{q}_1,\mathbf{q}_2,\ldots,\mathbf{q}_M} \mathbb{E}_{\mathbf{p}}\left\{ \left|\left| \mathbf{p}-\mathbf{q}_{\mathcal{E}(\mathbf{p})} \right|\right|^{\chi}\right\},
    \label{eqn:high_power_optztn}
\end{equation}
where $\chi$ is the power. This optimization problem can be solved by using a Lloyd-type algorithm that uses the distortion function 
\begin{equation}
    d_{\chi}\left(\mathbf{p},\underline{\mathbf{q}}\right) = \left|\left| \mathbf{p}-\mathbf{q}_{\mathcal{E}(\mathbf{p})} \right|\right|^{\chi}.
\end{equation}
While the NNC step uses $d_{\chi}$, the CC step is utilizes the steepest descent method, with the update equation
\begin{equation}
\mathbf{q}_m^{(j+1)} = \mathbf{q}_m^{(j)} - \delta \frac{ \partial }{ \partial \mathbf{q}_m^{(j)} } \int\limits_{\mathcal{C}_m} d_{\chi}\left(\mathbf{p},\mathbf{q}_m^{(j)}\right) f_{\mathbf{P}}(\mathbf{p}) \mathrm{d}\mathbf{p},
\label{eqn:stp_desc_high_power}
\end{equation}
for all $m$, where $j$ is the iteration index, $\delta$ is the step size, and the gradient expression is given by
\begin{multline}
    \frac{ \partial }{ \partial \mathbf{q}_m } \int\limits_{\mathcal{C}_m} d_{\chi}\left(\mathbf{p},\mathbf{q}_m\right) f_{\mathbf{P}}(\mathbf{p}) \mathrm{d}\mathbf{p}\\ = \frac{\chi}{\left| \mathcal{C}_m \right|} \sum\limits_{\mathbf{p}_k \in \mathcal{C}_m} \left( \mathbf{q}_m - \mathbf{p}_k \right) \left|\left| \mathbf{p}_k - \mathbf{q}_m \right|\right|^{\chi-2}.
\end{multline}
This Lloyd-type algorithm is called the \emph{Lloyd-$\chi$ algorithm} and the proof of the above result for gradient can be found in Appendix \ref{app:gradient_lloyd_chi_proof}. The algorithm is given in Algorithm \ref{alg:lloyd_chi}.
\begin{algorithm}
\caption{Lloyd-$\chi$ Algorithm}\label{alg:lloyd_chi}
\begin{algorithmic}[1]
\State Initialize random AP locations $\mathbf{q}_1^{(0)},\mathbf{q}_2^{(0)},\ldots,\mathbf{q}_M^{(0)}$.
\State Use the NNC to determine the cells $\mathcal{C}_1^{(i+1)},\mathcal{C}_2^{(i+1)},\ldots,\mathcal{C}_M^{(i+1)}$ such that
\begin{equation*}
\mathcal{C}_m^{(i+1)} \!=\! \left\{\! \mathbf{p}_k \!:\! d_{\chi}\!\left(\!\mathbf{p}_k,\mathbf{q}_m^{(i)}\!\right) \!\leq \! d_{\chi}\!\left(\!\mathbf{p}_k,\mathbf{q}_l^{(i)}\!\right)\!, \forall l \!\neq m \!\right\}.
\end{equation*}
\State Use the steepest descent method to determine the AP locations $\mathbf{q}_1^{(i+1)},\mathbf{q}_2^{(i+1)},\ldots,\mathbf{q}_M^{(i+1)}$ with the update equation
\begin{equation*}
    \mathbf{q}_m^{(j+1)} \!\!=\! \mathbf{q}_m^{(j)} -  \frac{\delta\chi}{\left| \mathcal{C}_m^{(i+1)} \right|} \!\!\sum\limits_{\mathbf{p}_k \in \mathcal{C}_m^{(i+1)}} \!\!\!\!\!\!( \mathbf{q}_m^{(j)} - \mathbf{p}_k )\! \left|\left| \mathbf{p}_k - \mathbf{q}_m^{(j)} \right|\right|^{\chi-2},
\end{equation*}
where $\mathbf{q}_m^{(i+1)} = \mathbf{q}_m^{(j+1)}$ after convergence.
\State Repeat from step 2 until convergence.
\end{algorithmic}
\end{algorithm}

The above formulations were developed by assuming a single user located at $\mathbf{p}$. However, in practice and according to the system model, $M$ APs serve $M$ users at the same time. Hence, we now consider the case where $M$ users are picked from the distribution.

\subsection{Multiple users case}

\subsubsection{Random user selection}

If the $M$ users are selected independently  from the overall distribution $f(\mathbf{p})$, then using an i.i.d. model the joint distribution of their positions can be represented as
\begin{equation}
    f_{\underline{{\mathbf{P}}}}(\underline{{\mathbf{p}}}) = \prod\limits_{m=1}^M f_{\mathbf{P}_m}(\mathbf{p}_m) = \prod\limits_{m=1}^M f(\mathbf{p}_m),
    \label{eqn:product_user_densities}
\end{equation}
where $\underline{\mathbf{p}} = \{ \mathbf{p}_1,\mathbf{p}_2,\ldots,\mathbf{p}_M\}$, the set of locations of the $M$ users. If we assume that the users do not interact with each other\footnote{This implies that a user does not influence the AP selection of any other user. In other words, the distortion function between a user at $\mathbf{p}_m$ and its closest AP $\mathbf{q}_{\mathcal{E}(\mathbf{p}_m)}$ is independent of the positions of all other users $\mathbf{p}_{m^\prime}$, where $m^\prime \neq m$. It is worth noting that multiple users can select the same AP as its closest one.}, then the objective function can be the sum of distortions incurred by each user with its closest AP, i.e., the optimization is of the form
\begin{equation}
    \argmin{\mathbf{q}_1,\mathbf{q}_2,\ldots,\mathbf{q}_M} \mathbb{E}_{\underline{\mathbf{p}}}\left\{ \sum\limits_{m=1}^M d(\mathbf{p}_m,\mathbf{q}_{\mathcal{E}(\mathbf{p}_m)}) \right\},
    \label{eqn:multiple_user_iid_opt}
\end{equation}
where $\mathbf{q}_{\mathcal{E}(\mathbf{p})}$ defined as before. The objective function in the above optimization can be simplified as
\begin{multline}
    \mathbb{E}_{\underline{\mathbf{p}}}\left\{ \sum\limits_{m=1}^M d(\mathbf{p}_m,\mathbf{q}_{\mathcal{E}(\mathbf{p}_m)}) \right\}\\
    \begin{aligned}
    &\overset{(a)}{=} \sum\limits_{m=1}^M \int d(\mathbf{p}_m,\mathbf{q}_{\mathcal{E}(\mathbf{p}_m)}) f(\mathbf{p}_m) d\mathbf{p}_m,\\
    &\overset{(b)}{=} M \cdot \int d(\mathbf{p},\mathbf{q}_{\mathcal{E}(\mathbf{p})}) f(\mathbf{p}) d\mathbf{p}, 
    \end{aligned}
\end{multline}
where $(a)$ is obtained by simplifying the expectation using \eqref{eqn:product_user_densities} and $(b)$ arises from the fact that each user is i.i.d. The final objective function thus is essentially the same as the single user case.

The above model is applicable in the following scenario. First, since the selection does not limit one user per small cell, the cells must be capable of dealing with more than one user with no multiple access interference. Secondly, since there is no ICI considered, each small cell must be assigned orthogonal resources.  This leads to an interesting resource allocation problem which we do not pursue in this work.

\subsubsection{Random selection of one user per cell without ICI}

The formulation described above considers $M$ users at a time, but fails to follow the system model as each user is not necessarily picked from the Voronoi region or cell in which its serving AP is present. Under this model, assuming again that the users at $\underline{\mathbf{p}} = \{ \mathbf{p}_1,\mathbf{p}_2,\ldots,\mathbf{p}_M\}$ do not interact with one another, the objective function to minimize would be the sum of the average distortion in each cell, i.e., the optimization is
\begin{equation}
    \argmin{\mathbf{q}_1,\mathbf{q}_2,\ldots,\mathbf{q}_M}  \sum\limits_{m=1}^M \mathbb{E}_{\underline{\mathbf{p}}}\left\{ d(\mathbf{p}_m,\mathbf{q}_m) \right\},
    \label{eqn:multiple_user_cell_opt}
\end{equation}
with the joint distribution of the user positions as
\begin{equation}
\label{eq:1userpercell}
    f(\underline{\mathbf{p}}) = \prod\limits_{m=1}^M f(\mathbf{p}_m | \mathbf{p}_m \in \mathcal{C}_m).
\end{equation}
The above objective function can be simplified as
\begin{equation}
    \begin{aligned}
    \sum\limits_{m=1}^M \mathbb{E}_{\underline{\mathbf{p}}}\left\{ d(\mathbf{p}_m,\mathbf{q}_m) \right\} 
    &= \sum\limits_{m = 1}^M \int\limits_{\mathbf{p} \in \mathcal{C}_m} d(\mathbf{p},\mathbf{q}_m) f(\mathbf{p} | \mathbf{p} \in \mathcal{C}_m) d\mathbf{p},\\
    &= \sum\limits_{m = 1}^M S_m,
    \end{aligned}
    \label{eqn:multiple_user_cell_obj}
\end{equation}
where $S_m$ is from \eqref{eqn:Sm_definition}. It is worth noting here the difference between this objective function and that of the Lloyd algorithm in \eqref{eqn:VQ_opt_func} where each term $S_m$ is weighted by the probability that the user is present in the cell $\Pr(\mathbf{p} \in \mathcal{C}_m)$.
The solution to the above objective function is then a Lloyd-type algorithm with the CC step unchanged, but with the NN step using weighted distortion functions, with the weights being the inverse of the proportion of users present in the cell. More specifically, the squared error distortion $d_{\text{SE}}(\mathbf{p}_m,\mathbf{q}_m)$ is pre-multiplied with a weight $w_m = 1/\Pr(\mathbf{p} \in \mathcal{C}_m) = K/N_m$, where $N_m$ is the number of users in $\mathcal{C}_m$. The NNC step is
\begin{equation}
	\mathcal{C}_m = \left\{ \mathbf{p} : w_m d_{\text{SE}}\left(\mathbf{p},\mathbf{q}_m\right) \leq w_l d_{\text{SE}}\left(\mathbf{p},\mathbf{q}_l\right), \forall l \neq m \right\}.
\end{equation}
We call this algorithm as the \emph{weighted MSE (WMSE) Lloyd algorithm}.
The proof of the above solution is provided in Appendix \ref{app:wmse_solution} and the algorithm is outlined in Algorithm \ref{alg:wmse_lloyd}.
\begin{algorithm}
\caption{WMSE Lloyd Algorithm}\label{alg:wmse_lloyd}
\begin{algorithmic}[1]
\State Initialize random AP locations $\mathbf{q}_1^{(0)},\mathbf{q}_2^{(0)},\ldots,\mathbf{q}_M^{(0)}$.
\State Use the NNC to determine the cells $\mathcal{C}_1^{(i+1)},\mathcal{C}_2^{(i+1)},\ldots,\mathcal{C}_M^{(i+1)}$ such that
\begin{equation*}
\mathcal{C}_m^{(i+1)} \!=\! \left\{\! \mathbf{p}_k \!\!:\! w_md_{\text{SE}}\!\left(\!\mathbf{p}_k,\mathbf{q}_m^{(i)}\!\right) \!\!\leq \!\! w_jd_{\text{SE}}\!\left(\!\mathbf{p}_k,\mathbf{q}_l^{(i)}\!\right)\!, \!\forall l \!\neq \!m \!\right\}.
\end{equation*}
\State Use the CC to determine the AP locations $\mathbf{q}_1^{(i+1)},\mathbf{q}_2^{(i+1)},\ldots,\mathbf{q}_M^{(i+1)}$ such that
	\begin{equation*}
	\mathbf{q}_m^{(i+1)} = \sum\limits_{\mathbf{p}_k \in \mathcal{C}_m^{(i+1)}} \mathbf{p}_k.
	\end{equation*}
\State Repeat from step 2 until convergence.
\end{algorithmic}
\end{algorithm}

\subsubsection{Random selection of one user per cell with ICI}

In all the above formulations, we have considered only SNR and the fact that users do not interact with one another. However, under the effects of ICI, users do interact with one another in the form of providing interfering signals at the APs which are serving the other users.
Thus, the distortion function between a user and its serving AP would be a function of all other users as well and under a similar fashion as in \eqref{eqn:multiple_user_cell_opt}, we can write the objective function as
\begin{multline}
\sum\limits_{m=1}^M \mathbb{E}_{\underline{\mathbf{p}}} \left\{ d(\mathbf{p}_m,\mathbf{q}_m, \underline{\mathbf{p}}_m^\prime)\right\}\\
= \sum\limits_{m = 1}^M \int\int\cdots\int d(\mathbf{p}_m,\mathbf{q}_m,\underline{\mathbf{p}}_m^\prime) f(\underline{\mathbf{p}}) d\underline{\mathbf{p}},
\label{eqn:multiple_user_ici_opt}
\end{multline}
where the (general) distortion function uses the term $\underline{\mathbf{p}}_m^\prime$ which denotes the set of user positions other than the user at $\mathbf{p}_m$ and $f(\underline{\mathbf{p}})$ is as in (\ref{eq:1userpercell})..
It is clear from the objective function in \eqref{eqn:multiple_user_ici_opt}, due to the dependency of the distortion function on the interfering users, the joint distribution $f(\underline{\mathbf{p}})$ cannot be simplified to consider each cell $\mathcal{C}_m$ independently as in \eqref{eqn:multiple_user_cell_obj}. This makes the said objective function difficult and intractable, and hence cannot be readily solved.

To deal with ICI in a tractable manner, we adopt a slightly different approach based on the following considerations.
Based on results obtained so far, VQ provides a good framework to solve throughput optimization problems by Lloyd-type algorithms, although without ICI. We have also seen that the optimization in \eqref{eqn:multiple_user_ici_opt}, which considers ICI, is difficult to solve and to derive an AP placement algorithm. Further, numerical simulations (shown in Experiment 1 of Section 
\ref{ssec:numerical_results}) show that the average achievable rate is very similar, whether the Lloyd or Lloyd-type algorithms described in this section are used.
Motivated by these three facts, in the next section, we show how the Lloyd algorithm can be modified to account for ICI in AP placement.

\section{Throughput Formulations Accounting for Inter-Cell Interference}
\label{sec:acct_for_ici}

To account for ICI in the VQ framework, we develop two distortion functions, namely, the interference and inter-AP distortion functions. 

\subsubsection{Interference distortion measure}

From \eqref{eqn:snr_argmin}, it is clear that the Lloyd algorithm maximizes only the desired signal component. In addition, we are now required to minimize the interference term.
To construct a distortion function that considers both the desired and interference signals, we consider the achievable per-user rate, as considered in Section \ref{sssec:single_user_rate}, but using the SINR expression from \eqref{eqn:sinr}.
For notational simplicity, this SINR can be rewritten using $T_{\text{SNR}}$ for the desired signal power in the numerator and $T_{\text{ICI}}$ for the interference signal power in the denominator as follows
\begin{equation}
    \phi_{k_{\mathcal{E}(\mathbf{p})}} = \frac{T_{\text{SNR}}}{1+T_{\text{ICI}}},
\end{equation}
where $T_{\text{SNR}} = \rho_r \beta_{\mathcal{E}(\mathbf{p})} |h_{\mathcal{E}(\mathbf{p})}|^2$ and $T_{\text{ICI}} = \rho_r \sum_{m^{\prime}\neq m} \beta_{m^{\prime}} |h_{m^{\prime}}|^2$. To recapitulate the notation, we use a single subscript for simplicity and while $h_{\mathcal{E}(\mathbf{p})}$ and $\beta_{\mathcal{E}(\mathbf{p})}$ are the small-scale and large-scale fading coefficients, respectively, for the user at $\mathbf{p}$ to the serving cell, $h_{m^\prime}$ and $\beta_{m^\prime}$ correspond to the same quantities for the same user, but to the non-serving AP $m^\prime$.
Approximating the rate with high SINR ($\phi_{k_{\mathcal{E}(\mathbf{p})}} \gg 1$) and $T_{\text{ICI}} \gg 1$, and simplifying, we get
\begin{equation}
    \log \phi_{k_{\mathcal{E}(\mathbf{p})}} \approx \log T_{\text{SNR}} + \log \frac{1}{T_{\text{ICI}}}.
    \label{eqn:rate_log_sum}
\end{equation}
It is worth nothing here that the log-sum inequality could be applied to separate the second term above as the sum of inverses of the individual ICI terms. Further, considering the above sum of logarithm terms, it is clear that the MM technique can be applied. However, finding a surrogate function in this case is not as straightforward as in the solution to the MM-Lloyd algorithm discussed in Section \ref{sssec:single_user_rate}. We believe that the insight obtained from \eqref{eqn:rate_log_sum} is sufficient to generate a solution for AP placement. To simplify further, we negate the quantity in \eqref{eqn:rate_log_sum} and approximate using the relation $\log x < x$ which yields
\begin{equation}
    -\log \phi_{k_{\mathcal{E}(\mathbf{p})}} < \frac{1}{T_{\text{SNR}}} + T_{\text{ICI}}.
    \label{eqn:rate_sum_log_ub}
\end{equation}
We have now expressed the negative of the rate as the sum of the powers of the inverse of the desired and interference terms. Therefore, to maximize rate or equivalently, minimize the negative of the rate, we need to maximize SNR and minimize ICI, corresponding to the first and second terms in \eqref{eqn:rate_sum_log_ub}, respectively. The equation presented also reveals the structure of the distortion function that we will use.

Accordingly, in line with the objective function for a Lloyd-type algorithm in \eqref{eqn:VQ_opt_func}, we average \eqref{eqn:rate_sum_log_ub} over the user positions and the random quantities, as before, to obtain:
\begin{equation}
    \mathbb{E}_{\mathcal{A},\mathcal{B},\underline{\mathbf{p}}}\left\{ \frac{1}{T_{\text{SNR}}} + T_{\text{ICI}} \right\} = \mathbb{E}_{\mathbf{p}} \left\{ \mathbb{E}_{\mathcal{A},\mathcal{B},\underline{\mathbf{p}}^\prime}\left\{ \frac{1}{T_{\text{SNR}}} +  T_{\text{ICI}} \right\} \right\},
    \label{eqn:expec_sum_log_ub}
\end{equation}
where we denote $\underline{\mathbf{p}}^\prime$ as the set of positions of the interfering users at $\mathbf{p}_{m^\prime}$ with respect to the user at $\mathbf{p}$, i.e., one user each from the cells $\mathcal{C}_{m^\prime}$, $m^\prime \neq \mathcal{E}(\mathbf{p})$, set $\mathcal{A} = \{h_{\mathcal{E}(\mathbf{p})},z_{\mathcal{E}(\mathbf{p})}\}$ defined as before, and set $\mathcal{B} = \{h_{m^\prime},z_{m^\prime}:m^\prime \neq \mathcal{E}(\mathbf{p})\}$ consists of the small-scale and shadow fading quantities for all interfering cells. We have also assumed that the served user position $\mathbf{p}$ is independent from the interfering user positions $\underline{\mathbf{p}}^\prime$.
Consequently, by carrying out the expectations in \eqref{eqn:expec_sum_log_ub} over $\mathcal{A}$, $\mathcal{B}$, and $\underline{\mathbf{p}}^\prime$, we can write the distortion function as
\begin{equation}
\begin{aligned}
d_{\text{IF}}\left(\mathbf{p},\underline{\mathbf{q}}\right) &= \kappa_1\left|\left| \mathbf{p} - \mathbf{q}_{\mathcal{E}(\mathbf{p})} \right|\right|^{\gamma}\\
&\phantom{=} + \kappa_2 \sum\limits_{m^{\prime} \neq \mathcal{E}(\mathbf{p})} \int\limits_{ \mathcal{C}_{m^{\prime}}} \frac{1}{ \left|\left| \mathbf{p}_{m^\prime}-\mathbf{q}_{\mathcal{E}(\mathbf{p})} \right|\right|^{\gamma} } f_{\underline{\mathbf{P}}^\prime}(\underline{\mathbf{p}}^{\prime}) \mathrm{d}\underline{\mathbf{p}}^\prime,\\
\end{aligned}
\end{equation}
where $f_{\underline{\mathbf{P}}^\prime}(\underline{\mathbf{p}}^{\prime})$ is the joint distribution of the locations of all the interfering users, $\kappa_1 = \mathbb{E}_{\mathcal{A}}\{1/\rho_r c_1 z_{\mathcal{E}(\mathbf{p})} |h_{\mathcal{E}(\mathbf{p})}|^2\}$, and $\kappa_2 = \mathbb{E}_{\mathcal{B}}\{\rho_r c_1 z_{m^\prime} |h_{m^\prime}|^2\}$. This is the \emph{interference distortion function} denoted by $d_{\text{IF}}$ and the corresponding Lloyd-type algorithm is called the \emph{Interference Lloyd algorithm}. For further simplification, we will assume that the distribution of users in each interfering cell is independent, leading to a simpler distortion measure
\begin{multline}
d_{\text{IF}}\left(\mathbf{p},\underline{\mathbf{q}}\right) = \left|\left| \mathbf{p} - \mathbf{q}_{\mathcal{E}(\mathbf{p})} \right|\right|^{\gamma}\\
+ \kappa \sum\limits_{m^{\prime} \neq \mathcal{E}(\mathbf{p})} \int\limits_{ \mathcal{C}_{m^{\prime}}} \frac{1}{ \left|\left| \mathbf{p}_{m^\prime}-\mathbf{q}_{\mathcal{E}(\mathbf{p})} \right|\right|^{\gamma} } f_{\mathbf{P}_{m^\prime}}(\mathbf{p}_{m^\prime}) \mathrm{d}\mathbf{p}_{m^\prime},\\
\label{eqn:intf_dist_expr}
\end{multline}
where $f_{\mathbf{P}_{m^\prime}}(\mathbf{p}_{m^\prime})$ is the distribution of the user in cell $\mathcal{C}_{m^{\prime}}$ and $\kappa \triangleq \kappa_2/\kappa_1$. We call $\kappa \geq 0$ as the \textit{trade-off factor} and determines the trade-off between desired signal and ICI power. $\kappa$ can be varied to determine the importance of ICI power over desired signal power.


To solve for the AP locations, the Interference Lloyd algorithm retains the NNC step and the steepest descent method is to be used for the CC step (update equation given in \eqref{eqn:stp_desc_high_power} above), both steps utilizing $d_{\text{IF}}$.
%
For the sake of implementation, the integral in $d_{\text{IF}}$ from \eqref{eqn:intf_dist_expr} is numerically approximated using the sample average over a large number of realizations of the user locations, and is written as
\begin{multline}
d_{\text{IF}}\left(\mathbf{p},\mathbf{q}_m\right) 
= \left|\left| \mathbf{p} - \mathbf{q}_{m} \right|\right|^{\gamma}\\
+ \kappa \sum\limits_{m^{\prime} \neq m} \frac{1}{\left| \mathcal{C}_{m^{\prime}} \right|} \sum\limits_{\mathbf{p}_{k_{m^\prime}} \in \mathcal{C}_{m^{\prime}}}      \frac{1}{ \left|\left| \mathbf{p}_{k_{m^\prime}}-\mathbf{q}_{m} \right|\right|^{\gamma} },
\label{eqn:interference_dist_soln}
\end{multline}
where $\mathbf{p}_{k_{m^\prime}}$ represents the $k^{\text{th}}$ realization of the user position in cell $\mathcal{C}_{m^\prime}$.
%
%
The gradient function in this update equation is 
	\begin{multline}
	\frac{ \partial }{ \partial \mathbf{q}_m } \left\{ \int\limits_{\mathcal{C}_m} d_{\text{IF}}\left(\mathbf{p},\mathbf{q}_m\right) f_{\mathbf{P}}(\mathbf{p}) \mathrm{d}\mathbf{p} \right\}\\
	\begin{aligned}
	&= \frac{\gamma}{\left| \mathcal{C}_m \right|} \sum\limits_{\mathbf{p}_k \in \mathcal{C}_m} \left( \mathbf{q}_m - \mathbf{p}_k \right) \left|\left| \mathbf{p}_k - \mathbf{q}_m \right|\right|^{\gamma-2}\\ 
	&\phantom{=}+ \kappa \sum\limits_{m^{\prime} \neq m} \frac{\gamma}{\left| \mathcal{C}_{m^{\prime}} \right|} \sum\limits_{\mathbf{p}_{k_{m^\prime}} \in \mathcal{C}_{m^{\prime}}}      \frac{\left( \mathbf{p}_{k_{m^\prime}}-\mathbf{q}_m \right)}{ \left|\left| \mathbf{p}_{k_{m^\prime}}-\mathbf{q}_m \right|\right|^{\gamma} }.
	\end{aligned}
	\label{eqn:gradient_intf_dist}
	\end{multline}
The proof of this result is given in Appendix \ref{app:res_interference_gradient} and the steps for this Lloyd-type algorithm are provided in Algorithm \ref{alg:interference_lloyd}.
\begin{algorithm}
\caption{Interference Lloyd Algorithm}\label{alg:interference_lloyd}
\begin{algorithmic}[1]
\State Initialize random AP locations $\mathbf{q}_1^{(0)},\mathbf{q}_2^{(0)},\ldots,\mathbf{q}_M^{(0)}$.
\State Use the NNC to determine the cells $\mathcal{C}_1^{(i+1)},\mathcal{C}_2^{(i+1)},\ldots,\mathcal{C}_M^{(i+1)}$ such that
	\begin{equation*}
	\mathcal{C}_m^{(i+1)} \!=\! \left\{\! \mathbf{p}_k \!:\! d_{\text{IF}}\!\left(\!\mathbf{p}_k,\mathbf{q}_m^{(i)}\!\right) \!\leq \! d_{\text{IF}}\!\left(\!\mathbf{p}_k,\mathbf{q}_l^{(i)}\!\right)\!, \forall l \!\neq m \!\right\}.
	\end{equation*}
\State Use the steepest descent method to determine the AP locations $\mathbf{q}_1^{(i+1)},\mathbf{q}_2^{(i+1)},\ldots,\mathbf{q}_M^{(i+1)}$ with the update equation
	\begin{multline*}
	\mathbf{q}_m^{(j+1)} 
	= \mathbf{q}_m^{(j)}\\
	\begin{aligned}
	&\phantom{=} - \delta \left( \frac{\gamma}{\left| \mathcal{C}_m^{(i+1)} \right|} \sum\limits_{\mathbf{p}_k \in \mathcal{C}_m^{(i+1)}} \left( \mathbf{q}_m^{(j)} - \mathbf{p}_k \right) \left|\left| \mathbf{p}_k - \mathbf{q}_m^{(j)} \right|\right|^{\gamma-2}\right.\\ 
	&\phantom{=} \left. + \kappa \sum\limits_{m^{\prime} \neq m} \frac{\gamma}{\left| \mathcal{C}_{m^{\prime}}^{(i+1)} \right|} \sum\limits_{\mathbf{p}_{k_{m^\prime}} \in \mathcal{C}_{m^{\prime}}^{(i+1)}}      \frac{\left( \mathbf{p}_{k_{m^\prime}}-\mathbf{q}_m^{(j)} \right)}{ \left|\left| \mathbf{p}_{k_{m^\prime}}-\mathbf{q}_m^{(j)} \right|\right|^{\gamma} } \right),
	\end{aligned}
	\end{multline*}
	which, after convergence, $\mathbf{q}_m^{(i+1)} = \mathbf{q}_m^{(j+1)}$.
\State Repeat from step 2 until convergence.
\end{algorithmic}
\end{algorithm}

\subsubsection{Inter-AP distortion measure}

\addtocounter{equation}{1}

Here, we develop an alternate distortion function that also accounts for ICI. Consider the interference distortion function $d_{\text{IF}}$ in \eqref{eqn:intf_dist_expr}. Each of the ICI terms in the summation in $d_{\text{IF}}$ can be approximated as follows
\begin{equation}
	\mathbb{E}_{\mathbf{p}_{m^\prime}}\left\{\frac{ 1 }{ \left|\left| \mathbf{p}_{m^\prime}-\mathbf{q}_{\mathcal{E}(\mathbf{p})} \right|\right|^{\gamma} }\right\} \approx \frac{1}{\left|\left| \mathbf{q}_{m^{\prime}}-\mathbf{q}_{\mathcal{E}(\mathbf{p})} \right|\right|^{\gamma}},
	\label{eqn:interAP_inequality}
\end{equation}
the justification of which is provided in Appendix \ref{app:interAP_ineq_proof}. 
Substituting \eqref{eqn:interAP_inequality}, we can simplify \eqref{eqn:intf_dist_expr} as
\begin{equation}
d_{\text{IA}}\left(\mathbf{p},\underline{\mathbf{q}}\right) = \left|\left| \mathbf{p} - \mathbf{q}_{\mathcal{E}(\mathbf{p})} \right|\right|^{\gamma} +  \kappa \sum\limits_{m^{\prime} \neq \mathcal{E}(\mathbf{p})} \frac{1}{ \left|\left| \mathbf{q}_{m^{\prime}}-\mathbf{q}_{\mathcal{E}(\mathbf{p})} \right|\right|^{\gamma} }.
\label{eqn:inta_dist_expr}
\end{equation}
We call $d_{\text{IA}}$ the \emph{inter-AP distortion measure} as the ICI term now involves the distances between the interfering APs and AP indexed by $\mathcal{E}(\mathbf{p})$. The corresponding Lloyd-type algorithm is called the \emph{Inter-AP Lloyd algorithm}.

The solution of the optimization problem using $d_{\text{IA}}$ is similar to that of the Interference Lloyd algorithm. For the steepest descent method, the gradient corresponding to $d_{\text{IA}}$ is given as
	\begin{multline}
	\frac{ \partial }{ \partial \mathbf{q}_m } \left\{ \int\limits_{\mathcal{C}_m} d_{\text{IA}}\left(\mathbf{p},\mathbf{q}_m\right) f_{\mathbf{P}}(\mathbf{p}) \mathrm{d}\mathbf{p} \right\}\\
	\begin{aligned}
	&= \frac{\gamma}{\left| \mathcal{C}_m \right|} \sum\limits_{\mathbf{p}_k \in \mathcal{C}_m} \left( \mathbf{q}_m - \mathbf{p}_k \right) \left|\left| \mathbf{p}_k - \mathbf{q}_m \right|\right|^{\gamma-2}\\ 
	&\phantom{=} + \kappa\gamma \sum\limits_{m^{\prime} \neq m} \frac{\mathbf{q}_{m^{\prime}}-\mathbf{q}_{m}}{ \left|\left| \mathbf{q}_{m^{\prime}}-\mathbf{q}_{m} \right|\right|^{\gamma+2} }.
	\end{aligned}
	\end{multline}
The proof of this result is omitted as it is similar in calculation to the gradient of the interference distortion function in \eqref{eqn:gradient_intf_dist} and the Inter-AP Lloyd algorithm is given in Algorithm \ref{alg:interAP_lloyd}.
\begin{algorithm}
\caption{Inter-AP Lloyd Algorithm}\label{alg:interAP_lloyd}
\begin{algorithmic}[1]
\State Initialize random AP locations $\mathbf{q}_1^{(0)},\mathbf{q}_2^{(0)},\ldots,\mathbf{q}_M^{(0)}$.
\State Use the NNC to determine the cells $\mathcal{C}_1^{(i+1)},\mathcal{C}_2^{(i+1)},\ldots,\mathcal{C}_M^{(i+1)}$ such that
	\begin{equation*}
	\mathcal{C}_m^{(i+1)} \!=\! \left\{\! \mathbf{p}_k \!:\! d_{\text{IA}}\!\left(\!\mathbf{p}_k,\mathbf{q}_m^{(i)}\!\right) \!\leq \! d_{\text{IA}}\!\left(\!\mathbf{p}_k,\mathbf{q}_l^{(i)}\!\right)\!, \forall l \!\neq m \!\right\}.
	\end{equation*}
\State Use the steepest descent method to determine the AP locations $\mathbf{q}_1^{(i+1)},\mathbf{q}_2^{(i+1)},\ldots,\mathbf{q}_M^{(i+1)}$ with the update equation
	\begin{multline*}
	\mathbf{q}_m^{(j+1)}
	= \mathbf{q}_m^{(j)}\\
	\begin{aligned}
	&\phantom{=} - \delta \left( \frac{\gamma}{\left| \mathcal{C}_m^{(i+1)} \right|} \sum\limits_{\mathbf{p}_k \in \mathcal{C}_m^{(i+1)}} \left( \mathbf{q}_m^{(j)} - \mathbf{p}_k \right) \left|\left| \mathbf{p}_k - \mathbf{q}_m^{(j)} \right|\right|^{\gamma-2}\right.\\ 
	&\phantom{=} \left. + \kappa \sum\limits_{m^{\prime} \neq m} \frac{\mathbf{q}_{m^{\prime}}^{(j)}-\mathbf{q}_{m}^{(j)}}{ \left|\left| \mathbf{q}_{m^{\prime}}^{(j)}-\mathbf{q}_{m}^{(j)} \right|\right|^{\gamma+2} } \right),
	\end{aligned}
	\end{multline*}
	which, after convergence, $\mathbf{q}_m^{(i+1)} = \mathbf{q}_m^{(j+1)}$.
\State Repeat from step 2 until convergence.
\end{algorithmic}
\end{algorithm}


Among the distortion functions discussed above, it is evident that the MSE distortion $d_{\text{SE}}$ has the lowest complexity. On observing the expressions for the interference $d_{\text{IF}}$ and inter-AP $d_{\text{IA}}$ distortions, we find that in the former, the summation for each interfering cell is over all of the users in that cell while in the latter, the net summation is only over interfering cells. Hence, $d_{\text{IA}}$ has lower implementation complexity than $d_{\text{IF}}$. We will also see in a later section that user association with $d_{\text{IA}}$ is relatively much simpler.

\section{Cell Association Strategies}
\label{sec:Cell Association}


In the previous sections, we have addressed the problem of how to place APs based on the user locations. For completeness, we now aim at answering the following two questions on cell association: \emph{When a new user enters the system, to which cell should it associate to? What metric should be used?} In this section, we elaborate on these two issues in the context of Lloyd and Lloyd-type algorithms. 
Accordingly, consider a user at location $\mathbf{p}_{\text{new}}$ that has entered the area after AP placement has already occurred and will associate to the AP at $\mathbf{q}_{m_{\text{new}}}$.

For the Lloyd and Lloyd-type algorithms developed in this paper, the user would associate to the AP that yields the lowest distortion value. This is a straightforward implementation of the NNC for each algorithm. Formally, if $d$ represents any of the distortion functions, $\mathbf{q}_{m_{\text{new}}}$ is determined as
\begin{equation}
\mathbf{q}_{m_{\text{new}}} = \left\{ \mathbf{q}_m : d\left(\mathbf{p}_{\text{new}},\mathbf{q}_m\right) \leq d\left(\mathbf{p}_{\text{new}},\mathbf{q}_l\right), \forall l \neq m \right\}.
\label{eqn:cell_assoc_ici}
\end{equation}
Clearly, for the WMSE Lloyd algorithm, we would use the weighted squared error distortion function. It is worth pointing out that since the distortion function in the Interference Lloyd algorithm involves summing over all users in other (interfering) cells, the complexity of such a calculation cannot be overlooked.
Instead, a cell association procedure \eqref{eqn:cell_assoc_ici} based on the simpler distortion measures of the Lloyd or the Inter-AP Lloyd algorithm can be undertaken as a low-complexity alternative. 
Note that the distortion function in the latter involves only the knowledge of the interfering APs positions. This is of greater practical value as opposed to knowing the positions of all interfering users in the Interference Lloyd algorithm. %
In summary, the Inter-AP Lloyd algorithm not only offers lower implementation complexity and thus a simpler cell association strategy, but is also of more practical value
compared to the Interference Lloyd algorithm.

\section{Simulation Methodology and Results}
\label{sec:Simulation Methodology}


\subsection{Simulation Parameters}

A geographical area of dimensions $2 \text{ km} \times 2 \text{ km}$ is considered, consisting of $M = 8$ APs and $K = 2000$ users, and one randomly selected user in each cell communicates with its associated AP. The pathloss model in \eqref{eqn:pathloss} is used with $\gamma = 2$, shadow fading $z_{mk}$ ignored as it is averaged out in Sections \ref{sec:Problem Formulations} and \ref{sec:acct_for_ici}, $c_0 = 75.86$ and $c_1 = 7.59 \times 10^{-7}$ as in \cite[eq. (4.36), eq. (4.37)]{nay18t} according to the COST 231 Hata propagation model, and $r_0 = 0.001$ km. Also, the value of the trade-off factor is chosen to be $\kappa = 5\times10^8$ and the step-size for the gradient descent is $\delta = 5\times10^{-5}$ for the Lloyd-$\chi$ algorithm and $\delta = 0.5$ for the ICI-aware Lloyd-type algorithms. Moreover, the uplink transmit power is  $\rho_r = 200$ mW and the user distribution is a Gaussian Mixture Model (GMM) of the form
\begin{equation}
f_{\mathbf{P}}(\mathbf{p}) = \sum\limits_{l=1}^L p_l\mathcal{N}\left(\mathbf{p}\vert\boldsymbol{\mu}_l,\sigma_l^2\mathbf{I}\right),
\end{equation}
where $\mathbf{I}$ is the identity matrix and $L$ is the number of mixture components, called \textit{groups} henceforth. For group $l$, $p_l$ is the mixture component weight, $\boldsymbol{\mu}_l$ is the mean, and $\sigma_l$ is the variance.
We set a user configuration with the parameters $L=3$, $\boldsymbol{\mu}_1 = [0.5, -0.5]^T$, $\boldsymbol{\mu}_2 = [0, 0.5]^T$, $\boldsymbol{\mu}_3 = [-0.5, 0]^T$, $\sigma_1 = \sigma_2 = \sigma_3 = 100$, $p_1=0.6$, and $p_2=p_3=0.2$.

\subsection{Performance Measures}

We use the per-user achievable rate of user $k_m$, which is calculated using SINR $\phi_{k_m}$ from \eqref{eqn:sinr}. As given in \cite[Ch. 4]{nay18t}, we can also write the achievable rate as
\begin{equation}
R_{k_m} = \mathbb{E}\left\{ \log_2\left(1 + \phi_{k_m}\right) \right\}
= \frac{1}{\ln 2} e^{\mu_k} \mathrm{Ei}\left(\mu_k\right),
\end{equation}
where
\begin{equation}
\mu_k = \frac{ 1 + \rho_r \sum\limits_{\substack{m^{\prime}=1\\m^{\prime}\neq m}}^M \beta_{mk_{m^{\prime}}} }{ \rho_r \beta_{mk_m} },
\end{equation}
and
\begin{equation}
\mathrm{Ei}\left(x\right) = \int\limits_{x}^{\infty} \frac{e^{-t}}{t} \mathrm{d}t,
\end{equation}
is the exponential integral.

For each of the proposed algorithms and the benchmark Lloyd algorithm, the maximum iteration number is set at $50$.
Each of the above performance measures is calculated through Monte Carlo simulations with $10,000$ iterations, choosing a set of users randomly for transmission each time. Cumulative distribution function (CDF) plots are generated for each measure, though normalized by the largest value so as to focus on the relative performance of the considered algorithms.
%
For comparison, we utilize the 95\%-likely metric that represents the best rate of the worst $5\%$ of the users (users closer to cell borders). We denote this by $R_{k_m}^{5\%}$.
To quantify the improvement in relative performance of the proposed algorithms over the Lloyd algorithm, we use the following measure expressed as percentage
\begin{equation}
\text{Improvement Ratio} = \frac{ R_{k_m}^{5\%,\text{Proposed}} - R_{k_m}^{5\%,\text{Lloyd}} }{ R_{k_m}^{5\%,\text{Lloyd}} } \times 100.
\end{equation}
All algorithms are initialized with the same initial AP locations for unbiased comparison.



\subsection{Numerical Results}
\label{ssec:numerical_results}

\begin{figure}[t!]
	\centering
	\includegraphics [scale=0.58] {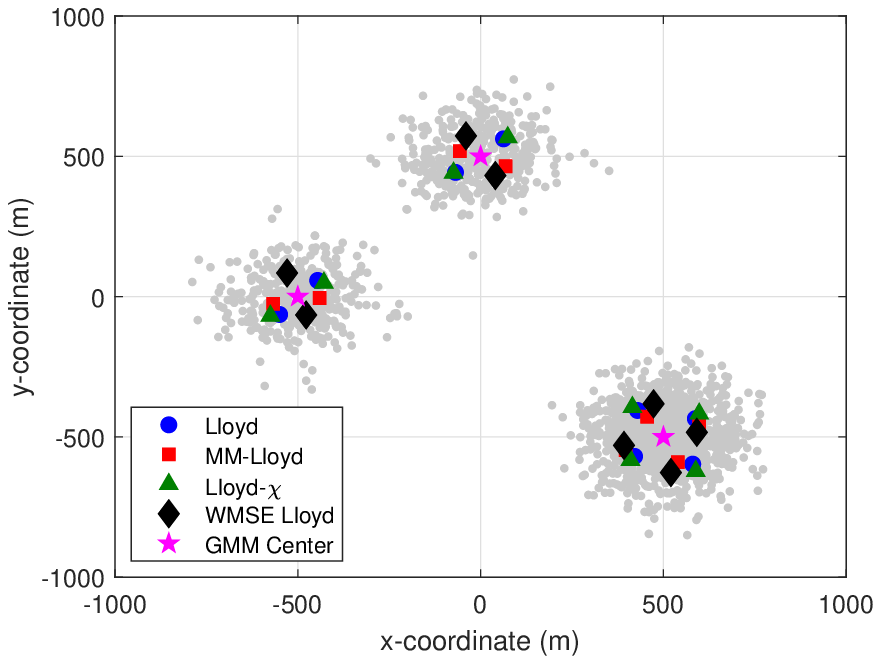}
	\caption{AP locations after convergence of the Lloyd, MM-Lloyd, Lloyd-$\chi$ ($\chi = 4$), and WMSE Lloyd algorithms with $M=8$.}
	\label{fig:scp_thrpt_motiv_final_locs_gmm1}
\end{figure}

\begin{figure}[t!]
	\centering
	\includegraphics [scale=0.58] {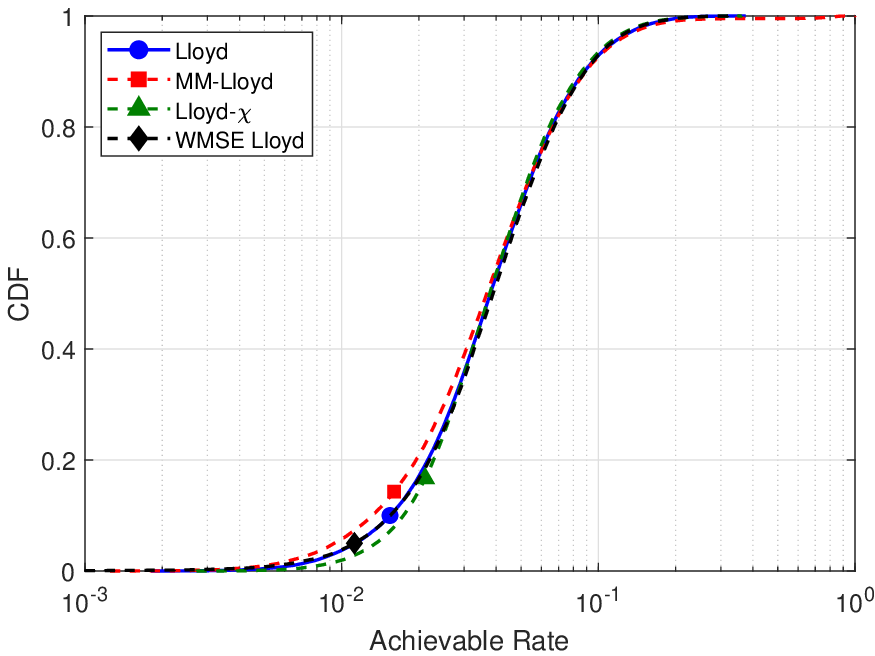}
	\caption{CDF plots of per-user achievable rate for the Lloyd, MM-Lloyd, Lloyd-$\chi$ ($\chi = 4$), and WMSE Lloyd algorithms with $M=8$.}
	\label{fig:scp_thrpt_motiv_ach_rate_norm_gmm1}
\end{figure}

\textit{Experiment 1.} We compare the throughput performances of the proposed Lloyd-type algorithms in Section \ref{sec:Problem Formulations} with the baseline Lloyd algorithm. For the Lloyd-$\chi$ algorithm, we use $\chi = 4$. The AP locations resultant from the algorithms are shown in Fig. \ref{fig:scp_thrpt_motiv_final_locs_gmm1}.
Relative to the AP positions of the Lloyd algorithm which are shown as blue circles, the APs in both the MM-Lloyd and WMSE Lloyd algorithms are placed closer to the GMM centers. For the MM-Lloyd algorithm, this can be explained by the logarithm in its objective function which suppresses the effect of users which are at large distances (e.g., cell periphery users away from the GMM center)  from the AP position during the placement process. This, in turn, causes the APs to position themselves closer to the GMM centers where the majority of the users at smaller distances are present. The WMSE Lloyd algorithm works in a different manner as the objective function in \eqref{eqn:multiple_user_cell_obj} is not weighted by the cell probabilities $\Pr(\mathbf{p} \in \mathcal{C}_m)$ as in \eqref{eqn:VQ_opt_func} of the Lloyd algorithm. This allows cells in the WMSE Lloyd algorithm to have a larger number of users than the Lloyd algorithm. On the other hand, the objective function of the Lloyd-$\chi$ algorithm amplifies the contribution of the users at large distances and results in the AP  moving away from the GMM center.
The effects of these placements are observed in their achievable rate plots in Fig. \ref{fig:scp_thrpt_motiv_ach_rate_norm_gmm1}. For both the MM-Lloyd and WMSE Lloyd algorithms, we observe that due to their AP positions, the lower rate suffers a reduction in comparison to the Lloyd algorithm. Nevertheless, note that there are more users achieving higher rates (right side of the CDF plot), particularly for the MM-Lloyd algorithm. The average rate values, however, are higher than that of the Lloyd algorithm, up to about $4\%$, as shown in Table \ref{tab:scp_thrpt_motiv_avg_rate}. On the other hand, the opposite of these effects are observed for the Lloyd-$\chi$ algorithm, with higher low rate values and lower average rate (only $0.65\%$ lower) than the Lloyd algorithm. Although omitted here due to space constraints, these effects increase as the power $\chi$ increases. 

\begin{table}[t!]
    \renewcommand{\arraystretch}{1.3}
	\centering 
	\caption{Percentage Improvements in Average Achievable Rates for the Lloyd-Type Algorithms of Section \ref{sec:Problem Formulations}}
	\label{tab:scp_thrpt_motiv_avg_rate}
	\begin{tabular}{|c|c|} 
		\hline
		Algorithm & Average Achievable Rate\\
		\hline
		MM-Lloyd & $4.26\%$ \\
		Lloyd-$\chi$ ($\chi = 4$) & $-0.64\%$ \\
		WMSE Lloyd & $1.14\%$ \\
		\hline
	\end{tabular}
\end{table}


\begin{figure}[t!]
	\centering
	\includegraphics [scale=0.58] {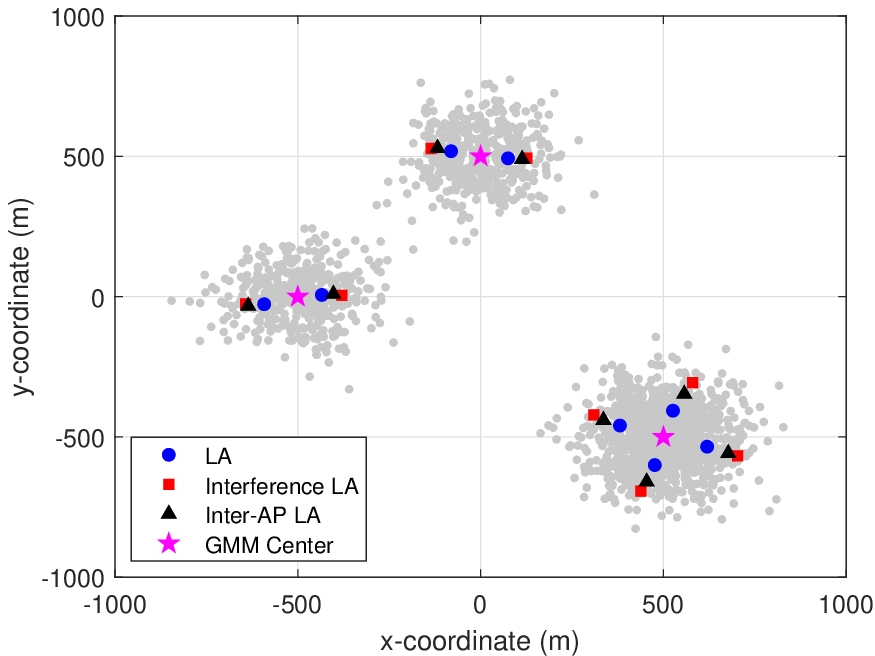}
	\caption{AP locations after convergence of the Lloyd and ICI-aware Lloyd-type algorithms with $\kappa = 5\times10^8$ and $M=8$.}
	\label{fig:scp_thrpt_final_locs_gmm1}
\end{figure}


\textit{Experiment 2.} Here, our simulations show throughput performances for the proposed ICI-aware Lloyd-type algorithms and the Lloyd algorithm, as well as their respective AP placements for comparison.   
%
%
The AP locations obtained after the algorithms converge are shown in Fig. \ref{fig:scp_thrpt_final_locs_gmm1}. AP locations for the Lloyd algorithm are shown as circles around the GMM center, which in turn are shown by stars. Compared to these positions, we can observe that the AP locations for both the Lloyd-type algorithms are situated further away from the GMM centers. For the Interference Lloyd algorithm, the AP positions denoted by the squares are the farthest. This is due to the interference term in its distortion function that forces neighboring cells apart. 
This effect is different (smaller) for the Inter-AP Lloyd algorithm due to the inter-AP distances term in its distortion function in contrast to the interference term in the Interference Lloyd algorithm.
%

\begin{figure}[t!]
	\centering
	\includegraphics [scale=0.58] {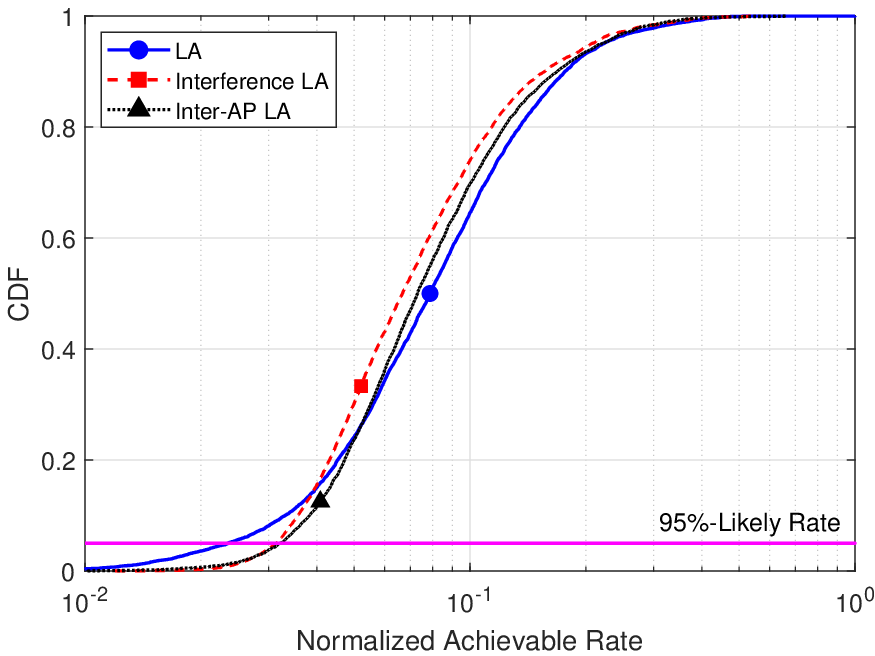}
	\caption{CDF plots of per-user achievable rate for the Lloyd and ICI-aware Lloyd-type algorithms with $\kappa = 5\times10^8$ and $M=8$.}
	\label{fig:scp_thrpt_ach_rate_gmm1}
\end{figure}

\begin{table}[t!]
    \renewcommand{\arraystretch}{1.3}
	\centering 
	\caption{Percentage Improvements in 95\%-Likely Achievable Rates for the ICI-Aware Lloyd-Type Algorithms}
	\label{tab:scp_thrpt_rate}
	\begin{tabular}{|c|c|} 
		\hline
		Algorithm & 95\%-Likely Achievable Rate\\
		\hline
		Interference Lloyd & $33.37\%$  \\
		Inter-AP Lloyd & $36.34\%$ \\
		\hline
	\end{tabular}
\end{table}

In Fig. \ref{fig:scp_thrpt_ach_rate_gmm1} we show the CDFs of the achievable rate obtained per user for each of the considered algorithms. The horizontal line at the 5$^{\text{th}}$ percentile shows the 95\%-likely rate and we compare the values where it intersects the throughput curves. It is clear that accounting for ICI during the AP placement procedure yields a superior performance to both Lloyd-type algorithms in comparison to the Lloyd algorithm in terms of the 95\%-likely rate. %
%
%
In practice, Fig. \ref{fig:scp_thrpt_ach_rate_gmm1} shows us that the worst $5\%$ of the users, usually the ones located closer to the cell borders and thus more susceptible to the deleterious effects of ICI, will have an uplink performance boost when APs are placed according to the proposed algorithms.
The percentage of improvements are quantified in Table \ref{tab:scp_thrpt_rate} from where we can confirm a very significant rate enhancement of up to $36.34\%$ in achievable rate, in comparison to the Lloyd algorithm. Also, from the same table, we can quantify that the Inter-AP Lloyd algorithm, despite its significantly lower computational complexity, performs slightly to moderately better than the Interference Lloyd algorithm, giving an approximately $3\%$ improvement in achievable rate. It is worth pointing out that in our experiments, lower $\kappa$ values resulted in less improvements as the Lloyd-type algorithms approached the results of the Lloyd algorithm. Higher $\kappa$ values resulted in convergence issues during the AP placement process. Many iterations of the algorithms were performed with other GMM configurations and $\kappa$ values. It was observed that the value of $\kappa = 5\times10^8$ was robust to a wide variety of user densities. Thus, the choice of $\kappa$ is an important part of the AP placement process and depends primarily on the area under consideration and the pathloss model. Finally, it is important to notice that although we have focused on the worst $5\%$ of the users, the Inter-AP Lloyd algorithm actually boosts the performances of the worst (nearly) $25\%$ of the users. The performance loss of the best users, as seen in the CDF plot, is justifiable due to the fact that users closer to the cell center tend to benefit from large SINR values that already suffice to provide them with more than their throughput requirements.

\section{Conclusion}
\label{sec:Conclusion}

In this paper, we have addressed the access point (AP) placement problem in the small-cell uplink paradigm under the criteria of throughput, while considering inter-cell interference (ICI).
After reviewing vector quantization (VQ), we explored related throughput formulations in the single user case and subsequently, the multiple user case corresponding to the small-cell model considered. Without ICI, we showed that the simple Lloyd algorithm performed similarly to the aforementioned formulations (only up to a $4\%$ difference) and could be a baseline algorithm to solve more complex problems.
Accordingly, we accounted for ICI in the optimization function of the Lloyd algorithm and mathematically arrived at two distinct distortion functions. 
Correspondingly, we proposed two Lloyd-type algorithms, namely the Interference Lloyd algorithm and the Inter-AP Lloyd algorithm. Both algorithms yield significant improvement to achievable rates, giving up to a marked $36.34\%$ increase in the 95\%-likely rate over the benchmark Lloyd algorithm. The Inter-AP Lloyd algorithm achieves throughput gains coupled with lower complexity and simpler user association over the Interference Lloyd algorithm. 
Finally, cell association strategies were outlined for all algorithms for completeness. 

\appendices

\section{Proof of Solution for MM-Lloyd Algorithm}
\label{app:mm_proof}

The expectation in the objective function in \eqref{eqn:rate_mm_opt} can be replaced by the sample average using the user realizations at $\mathbf{p}_k$ and written as
\begin{equation}
    J = \sum\limits_{\mathbf{p}_k \in \mathcal{C}_m} \log \left(||\mathbf{q}_m - \mathbf{p}_k||^2 + \epsilon\right),
    \label{eqn:rate_sample_obj}
\end{equation}
where $\mathbf{q}_{\mathcal{E}(\mathbf{p})}$ is replaced by $\mathbf{q}_m$ and the average is taken over all the users in cell $\mathcal{C}_m$ as the update steps correspond to the CC step of the Lloyd-type algorithm. Following the MM literature, a concave function can be upper bounded by its first-order Taylor expansion \cite{sun17}
\begin{equation}
    h(z) \leq h^\prime(z_l)(z - z_l) + h(z_l),
    \label{eqn:taylor_upper_bound}
\end{equation}
where $h(\cdot)$ is concave on $\mathbb{R}^+$, $z$ is the variable, $z_l$ is the point around which the expansion is carried out, and $h^\prime(\cdot)$ is the first derivative. In \eqref{eqn:rate_sample_obj}, we can take $h(z_k) = \log(z_k)$ and $z_k = ||\mathbf{q}_m - \mathbf{p}_k||^2 + \epsilon$, where we note that $z_k$ is scalar. Thus, using the upper bound \eqref{eqn:taylor_upper_bound} in \eqref{eqn:rate_sample_obj}, the objective function is
\begin{equation}
    J_1 = \sum\limits_{\mathbf{p}_k \in \mathcal{C}_m} h(z_k)
    \leq \sum\limits_{\mathbf{p}_k \in \mathcal{C}_m} [h^\prime(z_{k,l})(z_k - z_{k,l}) + h(z_{k,l})].
\end{equation}
Removing the terms that are not involved in the optimization, we have
\begin{equation}
\argmin{\mathbf{q}_m} \!\!\!\sum\limits_{\mathbf{p}_k \in \mathcal{C}_m} \!w_k z_k = \argmin{\mathbf{q}_m} \!\!\!\sum\limits_{\mathbf{p}_k \in \mathcal{C}_m} \!w_k \left(||\mathbf{q}_m - \mathbf{p}_k||^2 + \epsilon\right),
\label{eqn:rate_sample_upper_obj}
\end{equation}
where the weight $w_k$ is defined as
\begin{equation}
\begin{aligned}
    w_k &=  h^\prime(z_{k,l}) = \left.\frac{ \partial h(z_{k,l}) }{ \partial z_{k,l} } \right|_{z_{k,l} = ||\mathbf{q}_m - \mathbf{p}_k||^2 + \epsilon},\\
	&= \left.\frac{1}{z_{k,l}}\right|_{z_{k,l} =  ||\mathbf{q}_m - \mathbf{p}_k||^2 + \epsilon} = \frac{1}{||\mathbf{q}_m - \mathbf{p}_k||^2 + \epsilon},
	\end{aligned}
	\label{eqn:app_weight_upd}
\end{equation}
which gives the weight update equation. Now, given the weights, the objective function in \eqref{eqn:rate_sample_upper_obj} is
\begin{equation}
    J_2 = \sum\limits_{\mathbf{p}_k \in \mathcal{C}_m} w_k \left(||\mathbf{q}_m - \mathbf{p}_k||^2 + \epsilon\right).
\end{equation}
Taking the derivative and equating it to 0, i.e., $\partial J_2/\partial \mathbf{q}_m = 0$, gives the update equation for the AP position
\begin{equation}
	\mathbf{q}_m = \frac{ \sum\limits_{\mathbf{p}_k \in \mathcal{C}_m} w_k \mathbf{p}_k  }{ \sum\limits_{\mathbf{p}_k \in \mathcal{C}_m} w_k }.
\end{equation}

\section{Proof of Gradient for Lloyd-$\chi$ Algorithm}
\label{app:gradient_lloyd_chi_proof}

The gradient of the distortion function $d_{\chi}\left(\mathbf{p},\mathbf{q}_m\right)$ is calculated as
\begin{multline}
\frac{ \partial }{ \partial \mathbf{q}_m } \left\{ \int\limits_{\mathcal{C}_m} d_{\chi}\left(\mathbf{p},\mathbf{q}_m\right) f_{\mathbf{P}}(\mathbf{p}) \mathrm{d}\mathbf{p} \right\}\\
\begin{aligned}
&\overset{(a)}{=} \frac{ \partial }{ \partial \mathbf{q}_m } \left\{ \frac{1}{\left|\mathcal{C}_m\right|} \sum\limits_{\mathbf{p}_k \in \mathcal{C}_m} d_{\chi}\left(\mathbf{p}_k,\mathbf{q}_m\right) \right\},\\
&= \frac{ \partial }{ \partial \mathbf{q}_m } \left\{ \frac{1}{\left|\mathcal{C}_m\right|} \sum\limits_{\mathbf{p}_k \in \mathcal{C}_m} \left|\left| \mathbf{p}_k - \mathbf{q}_m \right|\right|^{\chi} \right\},\\
&\overset{(b)}{=} \frac{\chi}{\left| \mathcal{C}_m \right|} \sum\limits_{\mathbf{p}_k \in \mathcal{C}_m} \left( \mathbf{q}_m - \mathbf{p}_k \right) \left|\left| \mathbf{p}_k - \mathbf{q}_m \right|\right|^{\chi-2}.\\ 
\end{aligned}
\end{multline}
where $(a)$ is obtained by replacing the expectation with the sample mean and the factor of $2$ is assumed to be absorbed by the step-size $\delta$ in $(b)$.

\section{Proof of Solution for WMSE Lloyd Algorithm}
\label{app:wmse_solution}

Consider the simplified objective function in \eqref{eqn:multiple_user_cell_obj}, which can be written as
\begin{equation}
    \begin{aligned}
    \sum\limits_{m = 1}^M S_m 
    &\overset{(a)}{=} \sum\limits_{m = 1}^M \frac{1}{N_m} \sum\limits_{\mathbf{p}_k \in \mathcal{C}_m} d_{\text{SE}}(\mathbf{p}_k,\mathbf{q}_m),\\
    &\overset{(b)}{=} \frac{1}{K} \sum\limits_{m = 1}^M \frac{1}{\Pr(\mathbf{p} \in \mathcal{C}_m)} \sum\limits_{\mathbf{p}_k \in \mathcal{C}_m} d_{\text{SE}}(\mathbf{p}_k,\mathbf{q}_m),
    \end{aligned}
    \label{eqn:multiple_user_cell_obj_app}
\end{equation}
where in $(a)$, we have replaced the integral with the sample average over the users present in the cell and $N_m$ represents the number of users in cell $\mathcal{C}_m$, and in $(b)$, we have used $\Pr(\mathbf{p} \in \mathcal{C}_m) = N_m/K$, with $K$ as the total number of users. Comparing \eqref{eqn:multiple_user_cell_obj_app} with the objective function of the Lloyd algorithm $J_{\text{VQ}}$ in \eqref{eqn:VQ_opt_func}, we have
\begin{equation}
\begin{aligned}
\sum\limits_{m=1}^M S_m \Pr(\mathbf{p} \in \mathcal{C}_m) &= \sum\limits_{m = 1}^M \frac{1}{N_m} \sum\limits_{\mathbf{p}_k \in \mathcal{C}_m} d_{\text{SE}}(\mathbf{p}_k,\mathbf{q}_m) \times \frac{N_m}{K},\\
&= \frac{1}{K} \sum\limits_{m = 1}^M \sum\limits_{\mathbf{p}_k \in \mathcal{C}_m} d_{\text{SE}}(\mathbf{p}_k,\mathbf{q}_m),
\end{aligned}
\end{equation}
where we can observe that the objective function in \eqref{eqn:multiple_user_cell_obj_app} is a weighted MSE (WMSE) measure, with the weight related to AP $m$ given by $w_m = 1/\Pr(\mathbf{p} \in \mathcal{C}_m)$. Thus, the NNC step is updated to use a weighted squared error distortion function, i.e., $w_md_{\text{SE}}(\mathbf{p}_k,\mathbf{q}_m)$. The CC step however remains independent of the weights. This can be proven by taking the derivative of the objective function in \eqref{eqn:multiple_user_cell_obj_app} with respect to the AP location $\mathbf{q}_m$, which gives the AP location as
\begin{equation}
    \mathbf{q}_m = \sum\limits_{\mathbf{p}_k \in \mathcal{C}_m} \mathbf{p}_k.
\end{equation}

\section{Proof of Inequality in \eqref{eqn:interAP_inequality}}
\label{app:interAP_ineq_proof}

Consider the term in the denominator of \eqref{eqn:interAP_inequality}
\begin{equation}
|| \mathbf{p}_{m^\prime}-\mathbf{q}_{\mathcal{E}(\mathbf{p})} ||^2 = || \underbrace{\mathbf{p}_{m^\prime}-\mathbf{q}_{m^{\prime}}}_{\mathbf{y}}+\underbrace{\mathbf{q}_{m^{\prime}}-\mathbf{q}_{\mathcal{E}(\mathbf{p})}}_{\mathbf{x}} ||^2.
\label{eqn:csineq_xplusy}
\end{equation}
It is clear that the distance between the interfering user and its serving AP, denoted by $\mathbf{y}$, is smaller than the distance of that same AP from the nearest AP, denoted by $\mathbf{x}$, which means
\begin{equation}
\left|\left|\mathbf{p}_{m^\prime}-\mathbf{q}_{m^{\prime}}\right|\right| \leq \left|\left| \mathbf{q}_{m^{\prime}}-\mathbf{q}_{\mathcal{E}(\mathbf{p})} \right|\right| \Rightarrow ||\mathbf{y}|| \leq ||\mathbf{x}||.
\end{equation}
Given that $m^{\prime}$ indexes the interfering cells, we can classify these cells into cells that are the immediate neighbors of cell $\mathcal{C}_{\mathcal{E}(\mathbf{p})}$, denoted by $\mathcal{IN}(\mathcal{E}(\mathbf{p}))$ and those that are not, and are thus farther away. Hence, the two cases are
\begin{equation}
    \begin{aligned}
		\left|\left|\mathbf{x}\right|\right| &\geq \left|\left|\mathbf{y}\right|\right|, &\forall m^{\prime} \in \mathcal{IN}(\mathcal{E}(\mathbf{p})), \quad m^{\prime} \neq \mathcal{E}(\mathbf{p}),\\
		\left|\left|\mathbf{x}\right|\right| &\gg \left|\left|\mathbf{y}\right|\right|, &\forall m^{\prime} \notin \mathcal{IN}(\mathcal{E}(\mathbf{p})), \quad m^{\prime} \neq \mathcal{E}(\mathbf{p}).
	\end{aligned}
\end{equation}
However, to simplify, we make the optimistic assumption that $||\mathbf{x}|| \gg ||\mathbf{y}||$ holds true for all $m^{\prime} \neq \mathcal{E}(\mathbf{p})$. This gives
\begin{equation}
	\begin{aligned}
	||\mathbf{x} + \mathbf{y}||^2 &= ||\mathbf{x}||^2 + ||\mathbf{y}||^2 + 2||\mathbf{x}||||\mathbf{y}||\cos\theta,\\
	&= ||\mathbf{x}||^2 \left( 1 + \frac{||\mathbf{y}||^2}{||\mathbf{x}||^2} + \frac{||\mathbf{y}||\cos \theta}{||\mathbf{x}||} \right),\\
	&\approx ||\mathbf{x}||^2.
	\end{aligned}
\end{equation}%
Note that this relation holds true even when $\gamma$ assumes values other than $\gamma = 2$.
Thus, from \eqref{eqn:csineq_xplusy}, we have
\begin{equation}
	\mathbb{E}_{\mathbf{p}_{m^\prime}}\left\{\frac{ 1 }{ \left|\left| \mathbf{p}_{m^\prime}-\mathbf{q}_{\mathcal{E}(\mathbf{p})} \right|\right|^{\gamma} }\right\} \approx \frac{1}{\left|\left| \mathbf{q}_{m^{\prime}}-\mathbf{q}_{\mathcal{E}(\mathbf{p})} \right|\right|^{\gamma}}.
\end{equation}

\section{Proof of Gradient for Interference Lloyd Algorithm}
\label{app:res_interference_gradient}
The gradient is calculated using the distortion function as
\begin{multline}
\frac{ \partial }{ \partial \mathbf{q}_m } \left\{ \int\limits_{\mathcal{C}_m} d_{\text{IF}}\left(\mathbf{p},\mathbf{q}_m\right) f_{\mathbf{P}}(\mathbf{p}) \mathrm{d}\mathbf{p} \right\}\\
\begin{aligned}
&= \frac{ \partial }{ \partial \mathbf{q}_m } \left\{ \frac{1}{\left|\mathcal{C}_m\right|} \sum\limits_{\mathbf{p}_k \in \mathcal{C}_m} d_{\text{IF}}\left(\mathbf{p}_k,\mathbf{q}_m\right) \right\},\\
&= \frac{ \partial }{ \partial \mathbf{q}_m } \left\{ \frac{1}{\left|\mathcal{C}_m\right|} \sum\limits_{\mathbf{p}_k \in \mathcal{C}_m} \left|\left| \mathbf{p}_k - \mathbf{q}_m \right|\right|^{\gamma}   \phantom{\sum\limits_{\mathbf{p}_{k_{m^\prime}} \in \mathcal{C}_{m^{\prime}}}      \frac{1}{ \left|\left| \mathbf{p}_{k_{m^\prime}}-\mathbf{q}_m \right|\right|^{\gamma} }}\right.\\
&\phantom{=} \left. + \kappa \sum\limits_{m^{\prime} \neq m} \frac{1}{\left| \mathcal{C}_{m^{\prime}} \right|} \sum\limits_{\mathbf{p}_{k_{m^\prime}} \in \mathcal{C}_{m^{\prime}}}      \frac{1}{ \left|\left| \mathbf{p}_{k_{m^\prime}}-\mathbf{q}_m \right|\right|^{\gamma} } \right\},\\
&= \frac{\gamma}{\left| \mathcal{C}_m \right|} \sum\limits_{\mathbf{p}_k \in \mathcal{C}_m} \left( \mathbf{q}_m - \mathbf{p}_k \right) \left|\left| \mathbf{p}_k - \mathbf{q}_m \right|\right|^{\gamma-2}\\ 
&\phantom{=} + \kappa \sum\limits_{m^{\prime} \neq m} \frac{\gamma}{\left| \mathcal{C}_{m^{\prime}} \right|} \sum\limits_{\mathbf{p}_{k_{m^\prime}}\in \mathcal{C}_{m^{\prime}}}      \frac{\left( \mathbf{p}_{k_{m^\prime}}-\mathbf{q}_m \right)}{ \left|\left| \mathbf{p}_{k_{m^\prime}}-\mathbf{q}_m \right|\right|^{\gamma} },
\end{aligned}
\end{multline}
where the factor of $2$ is assumed to be absorbed by the step-size $\delta$ as in Appendix \ref{app:gradient_lloyd_chi_proof}.
\ifCLASSOPTIONcaptionsoff
  \newpage
\fi



%
%
\bibliographystyle{IEEEtran}
\bibliography{references}

%




\end{document}